\documentclass[letter]{aa}
\usepackage[utf8]{inputenc}
\usepackage{threeparttable,pifont}
\usepackage{natbib}

\usepackage{pstricks}
\usepackage{color}

\usepackage{multicol}

\newcommand{\numb}[1]{\textcolor{orange}{#1}}
\renewcommand{\numb}[1]{#1}  

\newcommand{\adam}{\texttt{ADAM}}

\newcommand{\mistral}{\texttt{Mistral}}

\newcommand{\Dens}{3.99\,$\pm$\,0.26}
\newcommand{\Diam}{226\,$\pm$\,5}
\newcommand{\Mass}{(24.1\,$\pm$\,3.2)\,$\times$\,$10^{18}$}

\newcommand{\sid}{g$\cdot$cm$^{-3}$}

\begin{document}

\title{{(16) Psyche: A mesosiderite-like asteroid?}\thanks{Based on observations made with 
      1) ESO Telescopes at the La Silla
      Paranal Observatory under programs
      086.C-0785 (PI Carry) and
      199.C-0074 (PI Vernazza); and
      2) the W. M. Keck Observatory, which is operated as a scientific
      partnership among the California Institute of Technology, the
      University of California and the National Aeronautics and Space
      Administration. The Observatory
      was made possible by the generous financial support of the W. M. Keck
      Foundation.}}

\titlerunning{(16) Psyche and its environment using SPHERE}

\author{
    M.~Viikinkoski\inst{\ref{tampere}}     \and 
    P.~Vernazza\inst{\ref{lam}}            \and 
    J.~Hanu{\v s}\inst{\ref{prague}}       \and 
    H.~Le Coroller\inst{\ref{lam}}         \and 
    K.~Tazhenova\inst{\ref{lam}}           \and 
    B.~Carry\inst{\ref{oca}}               \and 
    M.~Marsset\inst{\ref{qub}}             \and 
    A.~Drouard\inst{\ref{lam}}             \and 
    F.~Marchis\inst{\ref{seti}}            \and 
    R.~Fetick\inst{\ref{lam}}              \and 
    T.~Fusco\inst{\ref{lam}}               \and 
    J.~{\v D}urech\inst{\ref{prague}}      \and 
    M.~Birlan\inst{\ref{imcce}}            \and 
    J.~Berthier\inst{\ref{imcce}}          \and 
    P.~Bartczak\inst{\ref{poznan}}         \and 
    C.~Dumas\inst{\ref{tmt}}               \and 
    J.~Castillo-Rogez\inst{\ref{jpl}}      \and 
    F.~Cipriani\inst{\ref{estec}}          \and 
    F.~Colas\inst{\ref{imcce}}             \and 
    M.~Ferrais\inst{\ref{liege}}           \and 
    J.~Grice\inst{\ref{oca},\ref{ou}}      \and 
    E.~Jehin\inst{\ref{liege}}             \and 
    L.~Jorda\inst{\ref{lam}}               \and 
    M.~Kaasalainen\inst{\ref{tampere}}     \and 
    A.~Kryszczynska\inst{\ref{poznan}}     \and 
    P.~Lamy\inst{\ref{lamos}}                \and 
    A.~Marciniak\inst{\ref{poznan}}        \and 
    T.~Michalowski\inst{\ref{poznan}}      \and 
    P.~Michel\inst{\ref{oca}}              \and 
    M.~Pajuelo\inst{\ref{imcce},\ref{puc}} \and 
    E.~Podlewska-Gaca\inst{\ref{poznan},\ref{edyta}} \and 
    T.~Santana-Ros\inst{\ref{poznan}}      \and 
    P.~Tanga\inst{\ref{oca}}               \and 
    F.~Vachier\inst{\ref{imcce}}           \and 
    A.~Vigan\inst{\ref{lam}}               \and 
    B.~Warner\inst{\ref{warner}}           \and 
    O.~Witasse\inst{\ref{estec}}           \and 
    B.~Yang\inst{\ref{eso}}                     
}

   \institute{
     Laboratory of Mathematics, Tampere University of Technology, PO Box 553, 33101, Tampere, Finland
     \email{matti.viikinkoski@tut.fi}\label{tampere}
     \and 
     Aix Marseille Univ, CNRS, LAM, Laboratoire d'Astrophysique de Marseille, Marseille, France
     \label{lam}
     \and 
     Astronomical Institute, Faculty of Mathematics and Physics, Charles University, V~Hole{\v s}ovi{\v c}k{\'a}ch 2, 18000 Prague, Czech Republic
     \label{prague}
     \and 
     Universit\'e C{\^o}te d'Azur, Observatoire de la C{\^o}te d'Azur, CNRS, Laboratoire Lagrange, France
     \label{oca}
     \and 
     Astrophysics Research Centre, Queen's University Belfast, BT7 1NN, UK
     \label{qub}
     \and 
     SETI Institute, Carl Sagan Center, 189 Bernado Avenue, Mountain View CA 94043, USA 
     \label{seti}
     \and 
     IMCCE, Observatoire de Paris, PSL Research University, CNRS, Sorbonne Universit{\'e}s, UPMC Univ Paris 06, Univ. Lille, France
     \label{imcce}
     \and 
     Astronomical Observatory Institute, Faculty of Physics, A. Mickiewicz University, Sloneczna 36, 60-286 Poznan
     \label{poznan}
     \and 
     Thirty-Meter-Telescope, 100 West Walnut St, Suite 300, Pasadena, CA 91124, USA
     \label{tmt}
     \and 
     Jet Propulsion Laboratory, California Institute of Technology, 4800 Oak Grove Drive, Pasadena, CA 91109, USA
     \label{jpl}
     \and 
     European Space Agency, ESTEC - Scientific Support Office, Keplerlaan 1, Noordwijk 2200 AG, The Netherlands
     \label{estec}
     \and 
     Laboratoire Atmosph\`eres, Milieux et Observations Spatiales, CNRS \& 
    Universit\'e de Versailles Saint-Quentin-en-Yvelines, Guyancourt, France
    \label{lamos}
     \and 
     Space sciences, Technologies and Astrophysics Research Institute, Universit{\'e} de Li{\`e}ge, All{\'e}e du 6 Ao{\^u}t 17, 4000 Li{\`e}ge, Belgium
     \label{liege}
     \and 
     Open University, School of Physical Sciences, The Open University, MK7 6AA, UK
     \label{ou}
     \and 
     Secci{\'o}n F{\'i}sica, Departamento de Ciencias, Pontificia Universidad Cat{\'o}lica del Per{\'u}, Apartado 1761, Lima, Per{\'u}
     \label{puc}
     \and 
     Institute of Physics, University of Szczecin, Wielkopolska 15, 70-453 Szczecin, Poland
     \label{edyta}
     \and 
     Center for Solar System Studies, 446 Sycamore Ave., Eaton, CO 80615, USA
     \label{warner}
     \and 
     European Southern Observatory (ESO), Alonso de Cordova 3107, 1900 Casilla Vitacura, Santiago, Chile
     \label{eso}
}

\date{}
\abstract{ 
  Asteroid (16) Psyche is the target of the NASA Psyche mission.
  It is considered  one of the few main-belt bodies that
  could be an exposed proto-planetary metallic core and that
  would thus be related to iron meteorites. Such
  an association is however challenged by both its near- and 
  mid-infrared spectral properties and the reported estimates
  of its density.
}{ 
  Here, we aim to refine the density of (16) Psyche  to set further constraints on its bulk composition and determine its potential meteoritic analog. 
}{ 
  We observed (16) Psyche with ESO VLT/SPHERE/ZIMPOL as part of our large program (ID 199.C-0074). 
  We used the high angular resolution of these observations to refine Psyche's three-dimensional (3D) shape model and subsequently its density when combined with the most recent mass estimates. In addition, we searched for potential companions around the asteroid.
}{ 
  We derived a bulk density of \numb{\Dens}\,\sid{} for Psyche. While such density is incompatible at the \numb{3}-sigma level with any iron meteorites ($\sim$\numb{7.8}\,\sid{}), it appears fully consistent with that of stony-iron meteorites such as mesosiderites (density $\sim$\numb{4.25}\,\sid{}). In addition, we found no satellite in our images and set an upper limit on the diameter of any non-detected satellite of 
  \numb{1460\,$\pm$\,200}\,m at 150\,km from Psyche (0.2\%\,$\times$\,R$_{Hill}$, the Hill radius) and
  \numb{800\,$\pm$\,200\,m} at 2,000\,km (3\%\,$\times$\,$R_{Hill}$). 
}{ 
  Considering that the visible and near-infrared spectral properties of mesosiderites are similar to those of Psyche, there is merit to a long-published
initial hypothesis that Psyche could be a plausible candidate parent body for mesosiderites. 
}


\maketitle

\section{Introduction}

\indent Asteroid (16) Psyche, the target of the NASA Discovery mission Psyche \citep{Elkins-Tanton2017}, is one of the very few main-belt asteroids that exhibits a relatively high radar albedo \citep[0.42\,$\pm$\,0.1,][]{Shepard2010, Shepard2015} and shallow phase-polarization minimum \citep{Dollfus1977, Dollfus1979}, which imply that its surface/subsurface is metal-rich.
On such a basis, it has been defined as a metallic world \citep{Elkins-Tanton2017} and is considered  one of the few main-belt bodies that could be an exposed planetary metallic core and that could thus be related to iron meteorites. This association is however challenged by both its near- and mid-infrared spectral properties and its density. 

\indent Spectroscopic observations in the near-infrared range have revealed the presence of (i) a weak ($\sim$1\%) 0.9\,$\mu$m absorption band suggesting the presence of orthopyroxene on its surface \citep[e.g.,][]{Hardersen2005, Ockert-Bell2008, Sanchez2017} and (ii) a 3 $\mu$m absorption feature suggesting the presence of hydrated silicates on its surface \citep{Takir2017}. Spectroscopic observations with the Spitzer Space Telescope in the thermal infrared have provided additional evidence for the presence of fine-grained silicates at the surface of Psyche but have also detected the presence of a metallic bedrock \citep{Landsman2018} consistent with earlier radar observations \citep{Shepard2010}.
Finally, its thermal inertia is among the highest for an asteroid of this size \citep{2013-Icarus-226-Matter}.
At this stage, one cannot exclude that some of the silicates (especially the hydrated ones) on the surface of the asteroid could have an exogenous origin \citep{Avdellidou2018}. We note that the presence of exogenous dust has already been reported at the surfaces of a number of large main-belt asteroids, including Vesta \citep{Reddy2012} and Ceres \citep{Vernazza2017}. 

\indent What is more puzzling at this stage is the reported density for Psyche. Recent estimates of its density all fall in the 3.7-4.7 g/cm$^{3}$ range \citep{Hanus2017b, Shepard2017, Drummond2018}. Considering that large main-belt asteroids with a diameter greater than 200\,km tend to have minimal macroporosity \citep[$\leq$10\%,][]{Carry2012,2015-Ast4-Scheeres}, one may expect this to be the case for Psyche (D$\sim$225km), which would significantly complicate a direct association between this asteroid and iron meteorites. 


\indent In the present study, we present high-angular-resolution imaging observations of Psyche with ESO VLT/SPHERE/ZIMPOL that were performed as part of our large program (ID 199.C-0074; PI: P. Vernazza). We use these observations to (i) refine Psyche's 3D shape model, and subsequently its density when combined with the most recent mass estimates, (ii) place for the first time constraints on the surface topography of the northern hemisphere of Psyche, and (iii) search for potential companions. Finally, we discuss what could be the most likely meteorite analog to Psyche.

\section{Observations and data processing}
\label{sect2}

\begin{figure*}[!ht]
  \begin{center}
    \includegraphics[width=\textwidth]{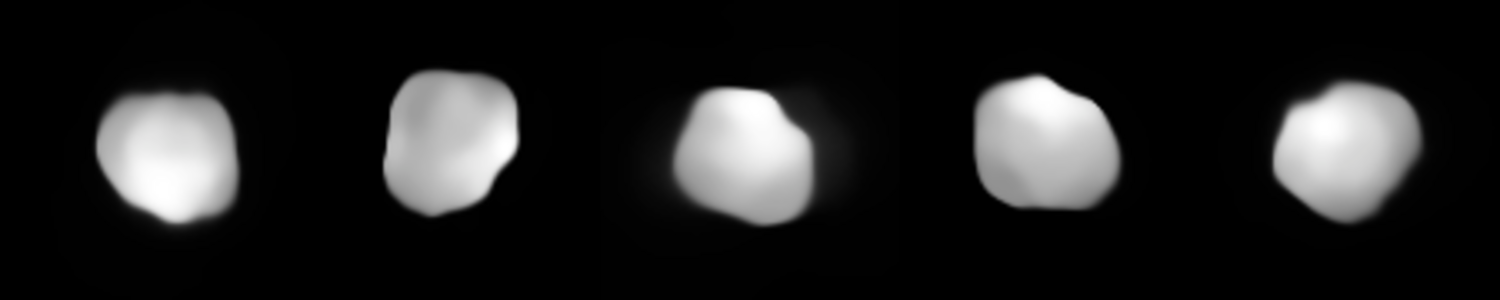}

    \vspace{-4pt}
    \includegraphics[width=\textwidth]{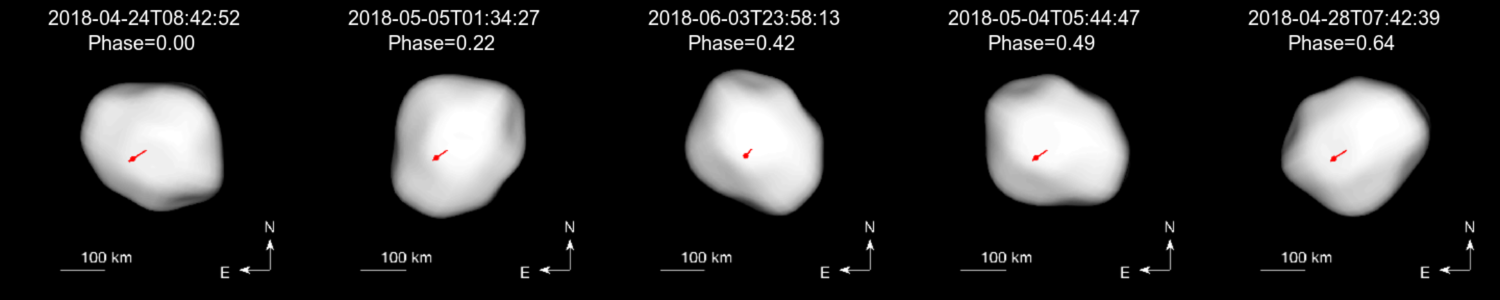}

    \vspace{-4pt}
    \includegraphics[width=\textwidth]{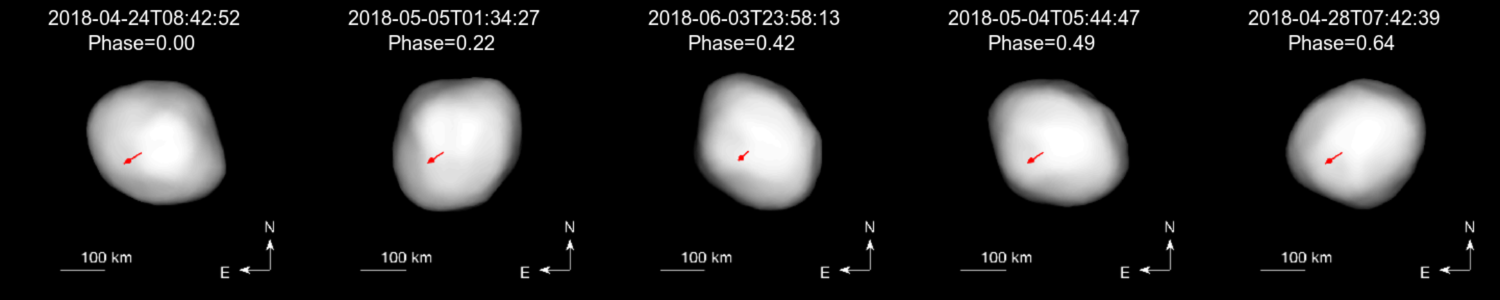}

     \vspace{-4pt}
    \includegraphics[width=\textwidth]{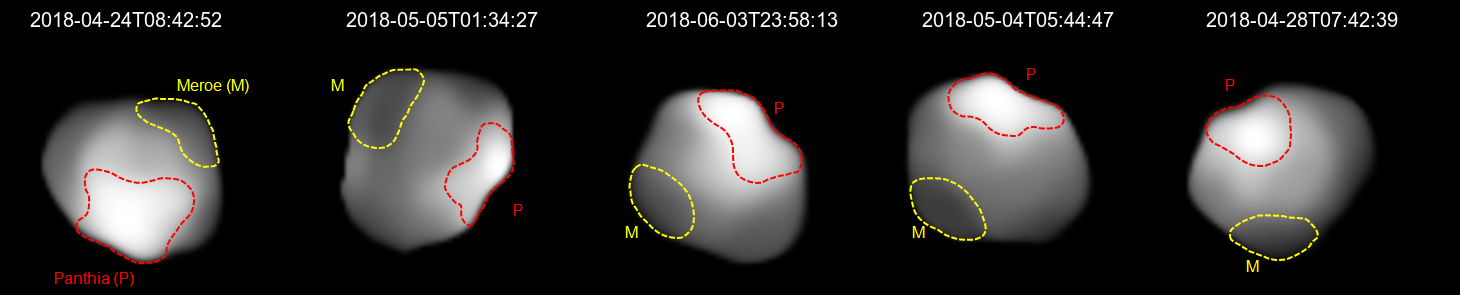}
  \end{center}
  \caption{\label{fig:comparison}
  \textbf{First row:} VLT/SPHERE/ZIMPOL images of Psyche obtained at five different epochs (ordered by rotation phase) and deconvolved with the \mistral{} algorithm and a parametric PSF with a Moffat shape (\ref{sect2}).
  \textbf{Second row:} Corresponding shape model projections.
  \textbf{Third row: } Corresponding projections using the shape model of \citet{Shepard2017}. The arrow shows the direction of the rotation axis.
  \textbf{Last row:} Images with contrast and size enhanced to highlight the albedo variegation. The two topographic regions that we consider as real features are also indicated.
}
\end{figure*}

\indent We observed Psyche with VLT/SPHERE/ZIMPOL \citep{Beuzit2008, Thalmann2008} at five different epochs, close to its opposition. Observational circumstances are listed in Table~\ref{tab:ao}. The data reduction was performed as described in \citet{2018-AA-Vernazza}. Finally, the optimal angular resolution of each image was restored with the \mistral{} algorithm \citep{Fusco2003, mugnier2004}.

\indent Instead of using a stellar point-spread function (PSF) for the deconvolution, we used for the first time a parametric PSF with a Moffat shape \citep{moffat1969}. During our large program, deconvolution with the stellar PSF acquired just after the asteroid observation has sometimes led to unsatisfactory results. On some occasions, using stellar PSFs acquired on different nights improved the results. 

\indent We therefore tested deconvolution with PSFs modeled by a 2D Moffat function. Whereas the results of this Moffat-PSF deconvolution were similar to those obtained with the stellar PSF, the deconvolution with a Moffat PSF always converged toward satisfactory results \citep{2018-AA-Fetick}.
We therefore started systematically using a parametric PSF with a Moffat shape to deconvolve our images. The deconvolved images of asteroid Psyche are shown in Fig.~\ref{fig:comparison} and Fig.~\ref{fig:DeconvAll}.

\indent Lastly, for each epoch we produced an average image to increase the signal-to-noise ratio (S/N) and to perform a search for satellites \citep[as we did for (41) Daphne, the first binary asteroid studied in our large program; see][]{2018-AA-Carry}.

\section{Moon search around Psyche}

\indent We did not discover any companion around Psyche. As a subsequent step, we estimated the minimum diameter above which a moon would have been detected. Defining a threshold S/N at which the detection occurs is complex because 1) the diffracted light (generating background photon noise) is not symmetric around Psyche, 2) the instrumental speckle noise does not follow a normal law, and 3) the sample size is small (small number of pixels (Moffats-PSF) at any small separation with
Psyche), thus preventing a robust statistical analysis. 

\indent To place an upper limit on the diameter of a companion, we  simulated \numb{ten} companions on a circle about on Airy disk away from the edge of the asteroid, where the photon noise from the diffracted light reaches its maximum value.
The artificial moons were represented by normalized Airy disks multiplied by a constant flux to reach a local S/N of 1--3, and by the typical Strehl ratio of our observations of \numb{0.7}. 
The flux of each moon ($F_m$) was then adjusted to be at the detection limit in each of our images. As such, a moon detectable around Psyche should have the following minimum diameter ($D_m$):

\begin{equation}
  D_m > D_{16} \times \left(\frac{F_m}{F_{16}}\right)^{(1/2)},
\end{equation} 

where $F_{16}$ is the integrated flux of Psyche and $D_{16}$ is the surface-equivalent diameter of Psyche on the plane of the sky (that is \numb{170}\,km). We found that any moon with a diameter $D_m$ above \numb{1460\,$\pm$\,200}\,m at 150 km from the primary (or 0.2\%\,$\times$\,R$_{Hill}$, the Hill radius) would have been detected.

We also explored more remote orbits typical of large main-belt asteroids \citep[at several primary radii; see][]{2015-AsteroidsIV-Margot}. At 100 pixels (about 2000\,km) from Psyche, corresponding to  3\%\,$\times$\,$R_{Hill}$ where most of the satellites of 100+km asteroids have been found \citep{2015-AsteroidsIV-Margot, yang2016}, the images are less affected by the diffracted light and we found that there should be no moon larger than \numb{$D_m$\,=\,800\,$\pm$\,200\,m} at these distances.


\section{Size, shape and density}

\begin{table*}
 \caption{\label{tab:param}Physical parameters of Psyche derived in this study compared with previous works. We list the volume-equivalent diameter $D$, dimensions along the major axes ($a$, $b$, $c$), sidereal rotation period $P$, pole ecliptic longitude $\lambda$ and latitude $\beta$, mass $m$, and bulk density $\rho$. The uncertainties are reported at 1\,$\sigma$.}
 \begin{center}
 \begin{threeparttable}
 \begin{tabular}{l r@{\,$\pm$\,}l r@{\,$\pm$\,}l r@{\,$\pm$\,}l r@{\,$\pm$\,}l r@{\,$\pm$\,}l}
  \hline
  Parameter & \multicolumn{2}{c}{K02/D11} & \multicolumn{2}{c}{S17} & \multicolumn{2}{c}{H17} & \multicolumn{2}{c}{D18} & \multicolumn{2}{c}{This work} \\
  \hline
   $D$ [km]             & 211 &  21 &   226 & 23   &  225 & 4     &  223 & 2     &  226 & 5   \\
   $\lambda$ [$^\circ$] &  32 &  5  &    34 &  5   &  28  & 4     &   32 & 3     &   34 & 3   \\
   $\beta$   [$^\circ$] & --7 &  5  &   --7 &  5   &  --6 & 3     &  --8 & 3     &   --9 & 3   \\
   $P$ & \multicolumn{2}{c}{4.195948(1)} & \multicolumn{2}{c}{4.195948(1)} & \multicolumn{2}{c}{4.195948(1)} & \multicolumn{2}{c}{4.195951(2)} & \multicolumn{2}{c}{4.195948(1)}   \\
   $a$ [km]             &    \multicolumn{2}{c}{}    &   279 & 27   &  293 & 5     &  274 & 2    &  290 & 5    \\
   $b$ [km]             &    \multicolumn{2}{c}{}    &   232 & 23   &  234 & 5     &  231 & 4    &  245 & 5    \\
   $c$ [km]             &    \multicolumn{2}{c}{}    &   189 & 19   &  167 & 8     &  176 & 3    &  170 & 8    \\
   $a/b$                &    \multicolumn{2}{c}{1.24}&  1.17 & 0.17 & 1.25 & 0.03  & 1.18 & 0.02 & 1.18 & 0.03 \\
   $b/c$                &    \multicolumn{2}{c}{1.27}&  1.21 & 0.17 & 1.40 & 0.07  & 1.31 & 0.03 & 1.44 & 0.07 \\
   $m$ [$10^{18}$ kg]   &    \multicolumn{2}{c}{}    &  27.2 & 7.5  & 22.3 & 3.6   & 24.3 & 3.5  & 24.1 & 3.2  \\
   $\rho$ [g/cm$^3$]    &    \multicolumn{2}{c}{}    &   4.5 & 1.4  &  3.7 & 0.6   & 4.16 & 0.64 & 3.99 & 0.26 \\
  \hline
 \end{tabular}
 \begin{tablenotes}[para,flushleft]
     \centering \footnotesize K02: \citet{Kaasalainen2002b} with diameter estimated from stellar occultation profiles by D11: \citet{Durech2011}, S17: \citet{Shepard2017}, H17: \citet{Hanus2017b}, D18: \citet{Drummond2018}.
 \end{tablenotes}
 \end{threeparttable}
 \end{center}
\end{table*}

 \indent Recently, \citet{Shepard2017} derived a shape model of Psyche using range-Doppler imaging. The S/N of radar echoes was sufficient for identifying two crater-like depressions in the southern hemisphere, but the northern hemisphere remained unobserved and modeled with an ellipsoidal shape. Our SPHERE observations are highly complementary because they covered the northern hemisphere.

\indent We used our All-Data Asteroid Modeling (\adam) inversion technique \citep{Viikinkoski2014, Viikinkoski2015, Viikinkoski2016, Hanus2017b} to reconstruct the 3D shape model and the spin of Psyche. We utilized all available  optical lightcurves (\numb{206}), stellar occultations (\numb{2}), and disk-resolved images (\numb{38}).
Our \numb{25} VLT/SPHERE/ZIMPOL images were complemented by \numb{11} Keck/NIRC2 and \numb{2} VLT/NACO imaging epochs \citep{Hanus2017b, Shepard2017, Drummond2018}, out of which \numb{8} were taken under different sub-observer latitudes, complementary to our SPHERE images (Table~\ref{tab:ao}).

\indent We also included two multiple-chord stellar occultations (Fig.~\ref{fig:occ}) recorded in 2010 and 2014 \citep{PDSSBN-OCC}. The 2014 occultation proved to be essential to constrain the size of the model along its rotation axis. Finally, we used a dataset of disk-integrated optical data consisting of \numb{209} lightcurves in total:
\numb{118} lightcurves from 
\citet{vanhouten1958}, 
\citet{chang1963}, 
\citet{Tedesco1985}, 
\citet{taylor1977}, 
\citet{Tedesco1983}, 
\citet{Lupishko1980}, 
\citet{Lupishko1982}, 
\citet{Zhou1981}, 
\citet{zhou1982}, 
\citet{weidenschilling1987}, 
\citet{harris1999}, 
\citet{Pfleiderer1987}, 
\citet{Dotto1992}, 
\citet{weidenschilling1990}, and 
\citet{neely1992}, 
already used by \citet{Hanus2017b}, 
\numb{3} lightcurves obtained by \citet{Warner2016} in 2015, 
\numb{85} lightcurves extracted from the SuperWASP image archive \citep{Grice2017}, and finally \numb{3} lightcurves obtained in 2018 by Emmanuel Jehin in the Rc filter using the TRAPPIST South telescope.

\indent The shape reconstruction was made more complex by discernible albedo variegation apparent both in the images and in the lightcurves. To model this phenomenon, we allowed each facet to have different relative brightness scaling parameters (i.e., to represent albedo variegation). 
Here, the albedo variegation was optimized simultaneously with the shape using both the lightcurves and imaging data, contrarily to \citet{Shepard2017} who used \numb{114} lightcurves to fit albedo variegation to the fixed shape model.
To avoid spurious spotty appearances, we defined a smoothing operator as a regularization term to discourage large deviations between neighboring facets. We chose, rather arbitrarily, $\pm 30\%$ as a reasonable interval for permitted variegation.
The albedo distribution on the model is based mainly on lightcurves and is therefore not unique. However, both LSL \citep{Kaasalainen2001b} and Hapke \citep{Hapke1984} scattering laws resulted in almost identical albedo distributions.

\indent We present in Fig.~\ref{fig:comparison} the five SPHERE epochs, together with the corresponding projections of our shape model and of the radar-derived shape model.
Our spin solution and volume-equivalent diameter $D$ (\numb{\Diam}\,km) agree well with those reported in \citet{Shepard2017} and in several other studies (Table~\ref{tab:param}).
The mass-deficit region \citep{Shepard2017} is clearly visible in each SPHERE image, but its shape and size differ from the radar estimates. Furthermore, the projected area of Psyche on the plane of the sky is systematically underestimated by the radar model, as suggested by the differences in diameter along the three axes reported in Table~\ref{tab:param}.
The polar region appears to be more flattened than suggested by the radar model.
The albedo distribution (Fig.~\ref{fig:model_albedo}) however closely matches  the one presented by \citet{Shepard2017}. 
The 3D shape model is depicted in Fig.~\ref{fig:model}. One should keep in mind that only features visible and consistent in at least two images can be considered as genuine. If a feature is only seen in one image its plausibility depends on how reliable one assumes the post-processing of the image to be.

\indent We then gathered \numb{27} mass estimates and \numb{15} diameter estimates of Psyche from the literature (Tables~\ref{tab:mass},\ref{tab:diam}), which show significant scatter (Fig.~\ref{fig:mass},\ref{fig:diam}). We therefore evaluated the reliability of each value and estimated a representative value of \numb{\Mass}\,kg following the procedure developed by \citet{Carry2012b} or \citet{2018-AA-Vernazza}. We used this value together with our new size estimate to compute Psyche's bulk density of \numb{\Dens}\,\sid{}.




\section{Surface topography}

\indent Although our observations covered only the northern hemisphere, they reveal the presence of two peculiar units with low and high brightness (Fig.~\ref{fig:comparison}). We named these two units after the twin witches related to the Metamorphoses, the latin novel of Apuleius in which Psyche is a character. 

\indent Of the two units, the dark unit \emph{Meroe} clearly appears as a crater. Using the same approach as the one described in \citet{2018-AA-Vernazza}, we measured its diameter to be in the 80-100\,km range. We subsequently estimated the width and depth of the depression present in the contour of the \emph{Panthia} unit to be $\sim$90\,km and $\sim$10\,km, respectively. Finally, the brightness profiles suggest that the \emph{Panthia} unit is about 7\% brighter than the surrounding areas, whereas the \emph{Meroe} unit is 8\% fainter.\\  

\section{Meteorite analogue}

\indent Whereas the density of Psyche derived in the present work (\numb{\Dens\,\sid{}}) is incompatible at the 3-sigma level with the one of iron meteorites ($\sim$7.8\,\sid{}), it appears fully consistent with that of stony-iron meteorites such as mesosiderites ($\sim$4.25\,\sid{}; \citealt{Britt2003}) or the Steinbach meteorite ($\sim$4.1\,\sid{}; \citealt{Britt2003}).

\indent Both mesosiderites and the unique Steinbach meteorite consist of a mixture of metal and pyroxene (orthopyroxene in the case of mesosiderites), a similar composition to that inferred for the surface of Psyche. Indeed, \citet{Vernazza2009} originally did not exclude a possible link between Psyche and mesosiderites on the basis of similar visible and near-infrared spectral properties, and recent laboratory measurements and observations \citep{Cloutis2017, Sanchez2017} reinforce this finding. 
Hence, there is merit to the initial hypothesis by \citet{Davis1999}, based on numerical simulations, that Psyche could be a plausible candidate parent body for mesosiderites. 



\begin{acknowledgements}
  \indent Some of the work presented here is
  based on observations collected 1) at the European Organization
  for Astronomical Research in the Southern Hemisphere under ESO
  programs
  086.C-0785 (PI Carry) and
  199.C-0074 (PI Vernazza); 2) by SuperWASP (DR1) 
  \citep{2010-AA-520-Butters} as provided by the WASP consortium, 
  and the computing and storage facilities at the CERIT Scientific
  Cloud, reg. no. CZ.1.05/3.2.00/08.0144 
  which is operated by Masaryk University, Czech Republic;
  and 3) the W.M. Keck Observatory, which is operated as a scientific partnership
  among the California Institute of Technology, the University of
  California and the National Aeronautics and Space
  Administration. 
  
  The 
  Observatory was made possible by the generous financial support
  of the W.M. Keck Foundation.
  This research has made use of the Keck Observatory Archive
  (KOA), which is operated by the W. M. Keck Observatory and the
  NASA Exoplanet Science Institute (NExScI), under contract with the
  National Aeronautics and Space Administration.\\
  \indent M. Viikinkoski and M. Kaasalainen were supported by the Academy of Finland Centre of Excellence in Inverse Problems. 
  B. Carry, A. Drouard, J. Grice and P. Vernazza were supported by
  CNRS/INSU/PNP. 
  J. Hanus and J. Durech were supported by the grant 18-09470S of the
  Czech Science Foundation. This work has been supported by Charles University Research program No. UNCE/SCI/023.
  F. Marchis was supported by the National Science Foundation under Grant No 1743015. The research leading to these results has received funding from the European Union's
 Horizon 2020 Research and Innovation Programme, under Grant Agreement no 687378.\\
  \indent The authors wish to recognize and acknowledge the very significant cultural role and reverence  that the summit of Mauna Kea
  has always had within the indigenous Hawaiian community. We
  are most fortunate to have the opportunity to conduct observations
  from this mountain.
  Thanks to all the amateurs worldwide who regularly observe
  asteroid lightcurves and stellar occultations. Co-authors of
  this study are amateurs who observed Psyche, and provided crucial data.\\
  \indent The authors acknowledge the use of the Virtual Observatory
  tools
  Miriade\,\footnote{Miriade: \url{http://vo.imcce.fr/webservices/miriade/}}
  \citep{2008-ACM-Berthier}, 
  TOPCAT\,\footnote{TOPCAT:
    \url{http://www.star.bris.ac.uk/~mbt/topcat/}}, and
  STILTS\,\footnote{STILTS: \url{http://www.star.bris.ac.uk/~mbt/stilts/}}
  \citep{2005-ASPC-Taylor}. This research used the
  SSOIS\,\footnote{SSOIS:
    \url{http://www.cadc-ccda.hia-iha.nrc-cnrc.gc.ca/en/ssois}}
  facility of the Canadian Astronomy Data Centre operated by the
  National Research Council of Canada with the support of the Canadian
  Space Agency \citep{2012-PASP-124-Gwyn}.
\end{acknowledgements}

\bibliography{all} 

\begin{thebibliography}{93}
\expandafter\ifx\csname natexlab\endcsname\relax\def\natexlab#1{#1}\fi

\bibitem[{{Avdellidou} {et~al.}(2018){Avdellidou}, {Delbo'}, \&
  {Fienga}}]{Avdellidou2018}
{Avdellidou}, C., {Delbo'}, M., \& {Fienga}, A. 2018, \mnras, 475, 3419

\bibitem[{{Baer} \& {Chesley}(2017)}]{2017-AJ-154-Baer}
{Baer}, J. \& {Chesley}, S.~R. 2017, \aj, 154, 76

\bibitem[{Baer {et~al.}(2011)Baer, Chesley, \& Matson}]{2011-AJ-141-Baer}
Baer, J., Chesley, S.~R., \& Matson, R.~D. 2011, \aj, 141, 143

\bibitem[{Baer {et~al.}(2008)Baer, {Milani}, Chesley, \&
  Matson}]{2008-DPS-40-Baer}
Baer, J., {Milani}, A., Chesley, S.~R., \& Matson, R.~D. 2008, in Bulletin of
  the American Astronomical Society, Vol.~40, 493

\bibitem[{Berthier {et~al.}(2008)Berthier, Hestroffer, {Carry}, {{\v D}urech},
  Tanga, {Delbo}, \& Vachier}]{2008-ACM-Berthier}
Berthier, J., Hestroffer, D., {Carry}, B., {et~al.} 2008, LPI Contributions,
  1405, 8374

\bibitem[{{Beuzit} {et~al.}(2008){Beuzit}, {Feldt}, {Dohlen}, {Mouillet},
  {Puget}, {Wildi}, {Abe}, {Antichi}, {Baruffolo}, {Baudoz}, {Boccaletti},
  {Carbillet}, {Charton}, {Claudi}, {Downing}, {Fabron}, {Feautrier},
  {Fedrigo}, {Fusco}, {Gach}, {Gratton}, {Henning}, {Hubin}, {Joos}, {Kasper},
  {Langlois}, {Lenzen}, {Moutou}, {Pavlov}, {Petit}, {Pragt}, {Rabou}, {Rigal},
  {Roelfsema}, {Rousset}, {Saisse}, {Schmid}, {Stadler}, {Thalmann}, {Turatto},
  {Udry}, {Vakili}, \& {Waters}}]{Beuzit2008}
{Beuzit}, J.-L., {Feldt}, M., {Dohlen}, K., {et~al.} 2008, in \procspie, Vol.
  7014, Ground-based and Airborne Instrumentation for Astronomy II, 701418

\bibitem[{{Britt} \& {Consolmagno}(2003)}]{Britt2003}
{Britt}, D.~T. \& {Consolmagno}, G.~J. 2003, Meteoritics and Planetary Science,
  38, 1161

\bibitem[{{Butters} {et~al.}(2010){Butters}, {West}, {Anderson}, {Collier
  Cameron}, {Clarkson}, {Enoch}, {Haswell}, {Hellier}, {Horne}, {Joshi},
  {Kane}, {Lister}, {Maxted}, {Parley}, {Pollacco}, {Smalley}, {Street},
  {Todd}, {Wheatley}, \& {Wilson}}]{2010-AA-520-Butters}
{Butters}, O.~W., {West}, R.~G., {Anderson}, D.~R., {et~al.} 2010, \aap, 520,
  L10

\bibitem[{{Carry}(2012)}]{Carry2012}
{Carry}, B. 2012, \planss, 73, 98

\bibitem[{{Carry} {et~al.}(2012){Carry}, {Kaasalainen}, {Merline},
  {M{\"u}ller}, {Jorda}, {Drummond}, {Berthier}, {O'Rourke}, {{\v D}urech},
  {K{\"u}ppers}, {Conrad}, {Tamblyn}, {Dumas}, {Sierks}, \& {OSIRIS
  Team}}]{Carry2012b}
{Carry}, B., {Kaasalainen}, M., {Merline}, W.~J., {et~al.} 2012, \planss, 66,
  200

\bibitem[{{Carry} {et~al.}(2018){Carry}, Vachier, Berthier, \&
  team}]{2018-AA-Carry}
{Carry}, B., Vachier, Berthier, \& team, L. 2018, \aap, submitted

\bibitem[{{Chang} \& S.(1963)}]{chang1963}
{Chang}, Y.~C. \& S., C.~C. 1963, Acta Astronomica Sinica, 11, 139

\bibitem[{{Cloutis} {et~al.}(2017){Cloutis}, {Applin}, {Kiddell}, {Tait},
  {Nicklin}, {DiCecco}, \& {Hardersen}}]{Cloutis2017}
{Cloutis}, E.~A., {Applin}, D.~M., {Kiddell}, C., {et~al.} 2017, in Lunar and
  Planetary Science Conference, Vol.~48, Lunar and Planetary Science
  Conference, 1228

\bibitem[{{Davis} {et~al.}(1999){Davis}, {Farinella}, \& {Marzari}}]{Davis1999}
{Davis}, D.~R., {Farinella}, P., \& {Marzari}, F. 1999, \icarus, 137, 140

\bibitem[{{Dollfus} {et~al.}(1979){Dollfus}, {Auriere}, \&
  {Santer}}]{Dollfus1979}
{Dollfus}, A., {Auriere}, M., \& {Santer}, R. 1979, \aj, 84, 1419

\bibitem[{{Dollfus} \& {Geake}(1977)}]{Dollfus1977}
{Dollfus}, A. \& {Geake}, J.~A. 1977, Philosophical Transactions of the Royal
  Society of London A: Mathematical, Physical and Engineering Sciences, 285,
  397

\bibitem[{{Dotto} {et~al.}(1992){Dotto}, {Barucci}, {Fulchignoni}, {di
  Martino}, {Rotundi}, {Burchi}, \& {di Paolantonio}}]{Dotto1992}
{Dotto}, E., {Barucci}, M.~A., {Fulchignoni}, M., {et~al.} 1992, \aaps, 95, 195

\bibitem[{{Drummond} \& {Christou}(2008)}]{Drummond2008}
{Drummond}, J. \& {Christou}, J. 2008, \icarus, 197, 480

\bibitem[{{Drummond} {et~al.}(2018){Drummond}, {Merline}, {Carry}, {Conrad},
  {Reddy}, {Tamblyn}, {Chapman}, {Enke}, {Pater}, {Kleer}, {Christou}, \&
  {Dumas}}]{Drummond2018}
{Drummond}, J.~D., {Merline}, W.~J., {Carry}, B., {et~al.} 2018, \icarus, 305,
  174

\bibitem[{Dunham {et~al.}(2017)Dunham, Herald, Frappa, Hayamizu, Talbot, \&
  Timerson}]{PDSSBN-OCC}
Dunham, D.~W., Herald, D., Frappa, E., {et~al.} 2017, {Asteroid Occultations},
  NASA Planetary Data System, {EAR-A-3-RDR-OCCULTATIONS-V15.0}

\bibitem[{{\v{D}urech} {et~al.}(2011){\v{D}urech}, {Kaasalainen}, {Herald},
  {Dunham}, {Timerson}, {Hanu\v{s}}, {Frappa}, {Talbot}, {Hayamizu}, {Warner},
  {Pilcher}, \& {Gal{\'a}d}}]{Durech2011}
{\v{D}urech}, J., {Kaasalainen}, M., {Herald}, D., {et~al.} 2011, Icarus, 214,
  652

\bibitem[{{Elkins-Tanton} {et~al.}(2017){Elkins-Tanton}, {Asphaug}, {Bell},
  {Bercovici}, {Bills}, {Binzel}, {Bottke}, {Brown}, {Goldsten}, {Jaumann},
  {Jun}, {Lawrence}, {Lord}, {Marchi}, {McCoy}, {Oh}, {Park}, {Peplowski},
  {Polanskey}, {Potter}, {Prettyman}, {Raymond}, {Russell}, {Scott}, {Stone},
  {Sukhatme}, {Warner}, {Weiss}, {Wenkert}, {Wieczorek}, {Williams}, \&
  {Zuber}}]{Elkins-Tanton2017}
{Elkins-Tanton}, L.~T., {Asphaug}, E., {Bell}, J.~F., {et~al.} 2017, in Lunar
  and Planetary Science Conference, Vol.~48, Lunar and Planetary Science
  Conference, 1718

\bibitem[{Fetick {et~al.}(2018)Fetick, Fusco, les, copains, du, \&
  LAM}]{2018-AA-Fetick}
Fetick, Fusco, les, {et~al.} 2018, \aap, in preparation

\bibitem[{{Fienga} {et~al.}(2011){Fienga}, {Laskar}, {Kuchynka}, {Manche},
  {Desvignes}, {Gastineau}, {Cognard}, \& {Theureau}}]{Fienga2011}
{Fienga}, A., {Laskar}, J., {Kuchynka}, P., {et~al.} 2011, Celestial Mechanics
  and Dynamical Astronomy, 111, 363

\bibitem[{{Fienga} {et~al.}(2009){Fienga}, {Laskar}, {Morley}, {Manche},
  {Kuchynka}, {Le Poncin-Lafitte}, {Budnik}, {Gastineau}, \&
  {Somenzi}}]{Fienga2009}
{Fienga}, A., {Laskar}, J., {Morley}, T., {et~al.} 2009, \aap, 507, 1675

\bibitem[{{Fienga} {et~al.}(2013){Fienga}, {Manche}, {Laskar}, {Gastineau}, \&
  {Verma}}]{2012-SciNote-Fienga}
{Fienga}, A., {Manche}, H., {Laskar}, J., {Gastineau}, M., \& {Verma}, A. 2013,
  ArXiv e-prints

\bibitem[{{Fienga} {et~al.}(2014){Fienga}, {Manche}, {Laskar}, {Gastineau}, \&
  {Verma}}]{2014-SciNote-Fienga}
{Fienga}, A., {Manche}, H., {Laskar}, J., {Gastineau}, M., \& {Verma}, A. 2014,
  Scientific notes

\bibitem[{Folkner {et~al.}(2009)Folkner, Williams, \&
  Boggs}]{2009-SciNote-Folkner}
Folkner, W.~M., Williams, J.~G., \& Boggs, D.~H. 2009, IPN Progress Report, 42,
  1

\bibitem[{{Fusco} {et~al.}(2003){Fusco}, {Mugnier}, {Conan}, {Marchis},
  {Chauvin}, {Rousset}, {Lagrange}, {Mouillet}, \& {Roddier}}]{Fusco2003}
{Fusco}, T., {Mugnier}, L.~M., {Conan}, J.-M., {et~al.} 2003, in \procspie,
  Vol. 4839, Adaptive Optical System Technologies II, ed. P.~L. {Wizinowich} \&
  D.~{Bonaccini}, 1065--1075

\bibitem[{{Goffin}(2014)}]{2014-AA-565-Goffin}
{Goffin}, E. 2014, \aap, 565, A56

\bibitem[{{Grice} {et~al.}(2017){Grice}, {Snodgrass}, {Green}, {Parley}, \&
  {Carry}}]{Grice2017}
{Grice}, J., {Snodgrass}, C., {Green}, S.~F., {Parley}, N.~R., \& {Carry}, B.
  2017, in Asteroids, Comets, and Meteors: ACM 2017

\bibitem[{Gwyn {et~al.}(2012)Gwyn, Hill, \& Kavelaars}]{2012-PASP-124-Gwyn}
Gwyn, S. D.~J., Hill, N., \& Kavelaars, J.~J. 2012, Publications of the
  Astronomical Society of the Pacific, 124, 579

\bibitem[{{Hanu\v{s}} {et~al.}(2013){Hanu\v{s}}, {Marchis}, \&
  {\v{D}urech}}]{Hanus2013b}
{Hanu\v{s}}, J., {Marchis}, F., \& {\v{D}urech}, J. 2013, \icarus, 226, 1045

\bibitem[{{Hanu\v{s}} {et~al.}(2017){Hanu\v{s}}, {Viikinkoski}, {Marchis},
  {\v{D}urech}, {Kaasalainen}, {Delbo'}, {Herald}, {Frappa}, {Hayamizu},
  {Kerr}, {Preston}, {Timerson}, {Dunham}, \& {Talbot}}]{Hanus2017b}
{Hanu\v{s}}, J., {Viikinkoski}, M., {Marchis}, F., {et~al.} 2017, \aap, 601,
  A114

\bibitem[{{Hapke}(1984)}]{Hapke1984}
{Hapke}, B. 1984, \icarus, 59, 41

\bibitem[{{Hardersen} {et~al.}(2005){Hardersen}, {Gaffey}, \&
  {Abell}}]{Hardersen2005}
{Hardersen}, P.~S., {Gaffey}, M.~J., \& {Abell}, P.~A. 2005, \icarus, 175, 141

\bibitem[{{Harris} {et~al.}(1999){Harris}, {Young}, {Bowell}, \&
  {Tholen}}]{harris1999}
{Harris}, A.~W., {Young}, J.~W., {Bowell}, E., \& {Tholen}, D.~J. 1999, Icarus,
  142, 173

\bibitem[{Ivantsov(2008)}]{2008-PSS-56-Ivantsov}
Ivantsov, A. 2008, \planss, 56, 1857

\bibitem[{{Kaasalainen} {et~al.}(2001){Kaasalainen}, {Torppa}, \&
  {Muinonen}}]{Kaasalainen2001b}
{Kaasalainen}, M., {Torppa}, J., \& {Muinonen}, K. 2001, Icarus, 153, 37

\bibitem[{{Kaasalainen} {et~al.}(2002){Kaasalainen}, {Torppa}, \&
  {Piironen}}]{Kaasalainen2002b}
{Kaasalainen}, M., {Torppa}, J., \& {Piironen}, J. 2002, Icarus, 159, 369

\bibitem[{Kochetova(2004)}]{2004-SoSyR-38-Kochetova}
Kochetova, O.~M. 2004, Solar System Research, 38, 66

\bibitem[{{Kochetova} \& {Chernetenko}(2014)}]{2014-SoSyR-48-Kochetova}
{Kochetova}, O.~M. \& {Chernetenko}, Y.~A. 2014, Solar System Research, 48, 295

\bibitem[{Konopliv {et~al.}(2011)Konopliv, Asmar, Folkner, Karatekin, Nunes,
  Smrekar, Yoder, \& Zuber}]{2011-Icarus-211-Konopliv}
Konopliv, A.~S., Asmar, S.~W., Folkner, W.~M., {et~al.} 2011, \icarus, 211, 401

\bibitem[{Krasinsky {et~al.}(2001)Krasinsky, Pitjeva, Vasiliev, \&
  Yagudina}]{2001-IAA-Krasinsky}
Krasinsky, G.~A., Pitjeva, E.~V., Vasiliev, M.~V., \& Yagudina, E.~I. 2001, in
  Communications of IAA of RAS

\bibitem[{{Kuchynka} \& {Folkner}(2013)}]{2013-Icarus-222-Kuchynka}
{Kuchynka}, P. \& {Folkner}, W.~M. 2013, \icarus, 222, 243

\bibitem[{{Kuzmanoski} \& {Kova{\v c}evi{\'c}}(2002)}]{Kuzmanoski2002}
{Kuzmanoski}, M. \& {Kova{\v c}evi{\'c}}, A. 2002, \aap, 395, L17

\bibitem[{{Landsman} {et~al.}(2018){Landsman}, {Emery}, {Campins}, {Hanu{\v
  s}}, {Lim}, \& {Cruikshank}}]{Landsman2018}
{Landsman}, Z.~A., {Emery}, J.~P., {Campins}, H., {et~al.} 2018, \icarus, 304,
  58

\bibitem[{{Lupishko} {et~al.}(1982){Lupishko}, {Belskaia}, {Tupieva}, \&
  {Chernova}}]{Lupishko1982}
{Lupishko}, D.~F., {Belskaia}, I.~N., {Tupieva}, F.~A., \& {Chernova}, G.~P.
  1982, Astronomicheskii Vestnik, 16, 101

\bibitem[{{Lupishko} {et~al.}(1980){Lupishko}, {Kiselev}, {Chernova}, \&
  {Belskaya}}]{Lupishko1980}
{Lupishko}, D.~F., {Kiselev}, N.~N., {Chernova}, G.~P., \& {Belskaya}, I.~N.
  1980, Pisma v Astronomicheskii Zhurnal, 6, 184

\bibitem[{{Margot} {et~al.}(2015){Margot}, {Pravec}, {Taylor}, {Carry}, \&
  {Jacobson}}]{2015-AsteroidsIV-Margot}
{Margot}, J.-L., {Pravec}, P., {Taylor}, P., {Carry}, B., \& {Jacobson}, S.
  2015, {Asteroid Systems: Binaries, Triples, and Pairs} (Univ. Arizona Press),
  355--374

\bibitem[{{Masiero} {et~al.}(2012){Masiero}, {Mainzer}, {Grav}, {Bauer},
  {Cutri}, {Nugent}, \& {Cabrera}}]{Masiero2012}
{Masiero}, J.~R., {Mainzer}, A.~K., {Grav}, T., {et~al.} 2012, \apjl, 759, L8

\bibitem[{{Matter} {et~al.}(2013){Matter}, {Delbo}, {Carry}, \&
  {Ligori}}]{2013-Icarus-226-Matter}
{Matter}, A., {Delbo}, M., {Carry}, B., \& {Ligori}, S. 2013, \icarus, 226, 419

\bibitem[{Moffat(1969)}]{moffat1969}
Moffat, A. 1969, Astronomy and Astrophysics, 3, 455

\bibitem[{{Morrison} \& {Zellner}(2007)}]{PDSSBN-TRIAD}
{Morrison}, D. \& {Zellner}, B. 2007, NASA Planetary Data System

\bibitem[{Mugnier {et~al.}(2004)Mugnier, Fusco, \& Conan}]{mugnier2004}
Mugnier, L.~M., Fusco, T., \& Conan, J.-M. 2004, JOSA A, 21, 1841

\bibitem[{{Neely}(1992)}]{neely1992}
{Neely}, A.~W. 1992, Minor Planet Bulletin, 19, 28

\bibitem[{{Ockert-Bell} {et~al.}(2008){Ockert-Bell}, {Clark}, {Shepard},
  {Rivkin}, {Binzel}, {Thomas}, {DeMeo}, {Bus}, \& {Shah}}]{Ockert-Bell2008}
{Ockert-Bell}, M.~E., {Clark}, B.~E., {Shepard}, M.~K., {et~al.} 2008, \icarus,
  195, 206

\bibitem[{{Pfleiderer} {et~al.}(1987){Pfleiderer}, {Pfleiderer}, \&
  {Hanslmeier}}]{Pfleiderer1987}
{Pfleiderer}, J., {Pfleiderer}, M., \& {Hanslmeier}, A. 1987, \aaps, 69, 117

\bibitem[{{Pitjeva}(2010)}]{2010-IAUS-261-Pitjeva}
{Pitjeva}, E.~V. 2010, in IAU Symposium, Vol. 261, IAU Symposium, ed. S.~A.
  {Klioner}, P.~K. {Seidelmann}, \& M.~H. {Soffel}, 170--178

\bibitem[{{Pitjeva}(2013)}]{2013-SoSyR-47-Pitjeva}
{Pitjeva}, E.~V. 2013, Solar System Research, 47, 386

\bibitem[{{Reddy} {et~al.}(2012){Reddy}, {Le Corre}, {O'Brien}, {Nathues},
  {Cloutis}, {Durda}, {Bottke}, {Bhatt}, {Nesvorny}, {Buczkowski}, {Scully},
  {Palmer}, {Sierks}, {Mann}, {Becker}, {Beck}, {Mittlefehldt}, {Li},
  {Gaskell}, {Russell}, {Gaffey}, {McSween}, {McCord}, {Combe}, \&
  {Blewett}}]{Reddy2012}
{Reddy}, V., {Le Corre}, L., {O'Brien}, D.~P., {et~al.} 2012, \icarus, 221, 544

\bibitem[{{Ryan} \& {Woodward}(2010)}]{2010-AJ-140-Ryan}
{Ryan}, E.~L. \& {Woodward}, C.~E. 2010, \aj, 140, 933

\bibitem[{{Sanchez} {et~al.}(2017){Sanchez}, {Reddy}, {Shepard}, {Thomas},
  {Cloutis}, {Takir}, {Conrad}, {Kiddell}, \& {Applin}}]{Sanchez2017}
{Sanchez}, J.~A., {Reddy}, V., {Shepard}, M.~K., {et~al.} 2017, \aj, 153, 29

\bibitem[{{Scheeres} {et~al.}(2015){Scheeres}, {Britt}, {Carry}, \&
  {Holsapple}}]{2015-Ast4-Scheeres}
{Scheeres}, D.~J., {Britt}, D., {Carry}, B., \& {Holsapple}, K.~A. 2015,
  {Asteroid Interiors and Morphology}, 745--766

\bibitem[{{Shepard} {et~al.}(2010){Shepard}, {Clark}, {Ockert-Bell}, {Nolan},
  {Howell}, {Magri}, {Giorgini}, {Benner}, {Ostro}, {Harris}, {Warner},
  {Stephens}, \& {Mueller}}]{Shepard2010}
{Shepard}, M.~K., {Clark}, B.~E., {Ockert-Bell}, M., {et~al.} 2010, \icarus,
  208, 221

\bibitem[{{Shepard} {et~al.}(2017){Shepard}, {Richardson}, {Taylor},
  {Rodriguez-Ford}, {Conrad}, {de Pater}, {Adamkovics}, {de Kleer}, {Males},
  {Morzinski}, {Close}, {Kaasalainen}, {Viikinkoski}, {Timerson}, {Reddy},
  {Magri}, {Nolan}, {Howell}, {Benner}, {Giorgini}, {Warner}, \&
  {Harris}}]{Shepard2017}
{Shepard}, M.~K., {Richardson}, J., {Taylor}, P.~A., {et~al.} 2017, \icarus,
  281, 388

\bibitem[{{Shepard} {et~al.}(2015){Shepard}, {Taylor}, {Nolan}, {Howell},
  {Springmann}, {Giorgini}, {Warner}, {Harris}, {Stephens}, {Merline},
  {Rivkin}, {Benner}, {Coley}, {Clark}, {Ockert-Bell}, \&
  {Magri}}]{Shepard2015}
{Shepard}, M.~K., {Taylor}, P.~A., {Nolan}, M.~C., {et~al.} 2015, \icarus, 245,
  38

\bibitem[{{Somenzi} {et~al.}(2010){Somenzi}, {Fienga}, {Laskar}, \&
  {Kuchynka}}]{Somenzi2010}
{Somenzi}, L., {Fienga}, A., {Laskar}, J., \& {Kuchynka}, P. 2010, \planss, 58,
  858

\bibitem[{{Sykes} {et~al.}(1998){Sykes}, {Brown}, \& {Raugh}}]{PDSSBN-IRAS}
{Sykes}, M., {Brown}, T., \& {Raugh}, A. 1998, NASA Planetary Data System

\bibitem[{{Takir} {et~al.}(2017){Takir}, {Reddy}, {Sanchez}, {Shepard}, \&
  {Emery}}]{Takir2017}
{Takir}, D., {Reddy}, V., {Sanchez}, J.~A., {Shepard}, M.~K., \& {Emery}, J.~P.
  2017, \aj, 153, 31

\bibitem[{{Taylor}(2005)}]{2005-ASPC-Taylor}
{Taylor}, M.~B. 2005, in Astronomical Society of the Pacific Conference Series,
  Vol. 347, Astronomical Data Analysis Software and Systems XIV, ed.
  P.~{Shopbell}, M.~{Britton}, \& R.~{Ebert}, 29

\bibitem[{{Taylor}(1977)}]{taylor1977}
{Taylor}, R.~C. 1977, \aj, 82, 441

\bibitem[{{Tedesco} \& {Taylor}(1985)}]{Tedesco1985}
{Tedesco}, E.~F. \& {Taylor}, R.~C. 1985, \icarus, 61, 241

\bibitem[{{Tedesco} {et~al.}(1983){Tedesco}, {Taylor}, {Drummond}, {Harwood},
  {Nickoloff}, {Scaltriti}, \& {Zappala}}]{Tedesco1983}
{Tedesco}, E.~F., {Taylor}, R.~C., {Drummond}, J., {et~al.} 1983, \icarus, 54,
  30

\bibitem[{{Thalmann} {et~al.}(2008){Thalmann}, {Schmid}, {Boccaletti},
  {Mouillet}, {Dohlen}, {Roelfsema}, {Carbillet}, {Gisler}, {Beuzit}, {Feldt},
  {Gratton}, {Joos}, {Keller}, {Kragt}, {Pragt}, {Puget}, {Rigal}, {Snik},
  {Waters}, \& {Wildi}}]{Thalmann2008}
{Thalmann}, C., {Schmid}, H.~M., {Boccaletti}, A., {et~al.} 2008, in \procspie,
  Vol. 7014, Ground-based and Airborne Instrumentation for Astronomy II, 70143F

\bibitem[{{Usui} {et~al.}(2011){Usui}, {Kuroda}, {M{\"u}ller}, {Hasagawa},
  {Ishiguro}, {Ootsubo}, {Ishihara}, {Kataza}, {Takita}, {Oyabu}, {Ueno},
  {Matsuhara}, \& {Onaka}}]{Usui2011}
{Usui}, F., {Kuroda}, D., {M{\"u}ller}, T.~G., {et~al.} 2011, \pasj, 63, 1117

\bibitem[{{van Houten-Groeneveld} \& {van Houten}(1958)}]{vanhouten1958}
{van Houten-Groeneveld}, I. \& {van Houten}, C.~J. 1958, \apj, 127, 253

\bibitem[{Vasiliev \& Yagudina(1999)}]{1999-IAA-Vasiliev}
Vasiliev, M.~V. \& Yagudina, E.~I. 1999, in Communications of IAA of RAS

\bibitem[{{Vernazza} {et~al.}(2009){Vernazza}, {Brunetto}, {Binzel}, {Perron},
  {Fulvio}, {Strazzulla}, \& {Fulchignoni}}]{Vernazza2009}
{Vernazza}, P., {Brunetto}, R., {Binzel}, R.~P., {et~al.} 2009, \icarus, 202,
  477

\bibitem[{{Vernazza} {et~al.}(2017){Vernazza}, {Castillo-Rogez}, {Beck},
  {Emery}, {Brunetto}, {Delbo}, {Marsset}, {Marchis}, {Groussin}, {Zanda},
  {Lamy}, {Jorda}, {Mousis}, {Delsanti}, {Djouadi}, {Dionnet}, {Borondics}, \&
  {Carry}}]{Vernazza2017}
{Vernazza}, P., {Castillo-Rogez}, J., {Beck}, P., {et~al.} 2017, \aj, 153, 72

\bibitem[{Vernazza {et~al.}(2018)Vernazza, Jorda, Hanu{\v s}, Carry, \&
  et~al.}]{2018-AA-Vernazza}
Vernazza, P., Jorda, L., Hanu{\v s}, J., Carry, B., \& et~al. 2018, \aap, in
  press

\bibitem[{{Viateau}(2000)}]{Viateau2000}
{Viateau}, B. 2000, \aap, 354, 725

\bibitem[{{Viikinkoski}(2016)}]{Viikinkoski2016}
{Viikinkoski}, M. 2016, PhD thesis, Tampere University of Technology

\bibitem[{Viikinkoski \& Kaasalainen(2014)}]{Viikinkoski2014}
Viikinkoski, M. \& Kaasalainen, M. 2014, Inverse Problems and Imaging, 8, 885

\bibitem[{{Viikinkoski} {et~al.}(2015){Viikinkoski}, {Kaasalainen}, \&
  {\v{D}urech}}]{Viikinkoski2015}
{Viikinkoski}, M., {Kaasalainen}, M., \& {\v{D}urech}, J. 2015, \aap, 576, A8

\bibitem[{{Viswanathan} {et~al.}(2017){Viswanathan}, {Fienga}, {Gastineau}, \&
  {Laskar}}]{2017-BDL-108-Viswanathan}
{Viswanathan}, V., {Fienga}, A., {Gastineau}, M., \& {Laskar}, J. 2017, Notes
  Scientifiques et Techniques de l'Institut de m{\'e}canique c{\'e}leste, (ISSN
  1621-3823), \#108, ISBN 2-910015-79-3 , 2017, 39 pp., 108

\bibitem[{{Warner}(2016)}]{Warner2016}
{Warner}, B.~D. 2016, Minor Planet Bulletin, 43, 137

\bibitem[{{Weidenschilling} {et~al.}(1990){Weidenschilling}, {Chapman},
  {Davis}, {Greenberg}, \& {Levy}}]{weidenschilling1990}
{Weidenschilling}, S.~J., {Chapman}, C.~R., {Davis}, D.~R., {Greenberg}, R., \&
  {Levy}, D.~H. 1990, Icarus, 86, 402

\bibitem[{{Weidenschilling} {et~al.}(1987){Weidenschilling}, {Chapman},
  {Davis}, {Greenberg}, {Levy}, \& {Vail}}]{weidenschilling1987}
{Weidenschilling}, S.~J., {Chapman}, C.~R., {Davis}, D.~R., {et~al.} 1987,
  \icarus, 70, 191

\bibitem[{{Yang} {et~al.}(2016){Yang}, {Wahhaj}, {Beauvalet}, {Marchis},
  {Dumas}, {Marsset}, {Nielsen}, \& {Vachier}}]{yang2016}
{Yang}, B., {Wahhaj}, Z., {Beauvalet}, L., {et~al.} 2016, \apjl, 820, L35

\bibitem[{{Zhou} \& {Yang}(1981)}]{Zhou1981}
{Zhou}, X.-H. \& {Yang}, X.-Y. 1981, Acta Astronomica Sinica, 22, 378

\bibitem[{{Zhou} {et~al.}(1982){Zhou}, {Yang}, \& {Wu}}]{zhou1982}
{Zhou}, X.~H., {Yang}, X.~Y., \& {Wu}, Z.~X. 1982, Acta Astronomica Sinica, 23,
  349

\bibitem[{Zielenbach(2011)}]{2011-AJ-142-Zielenbach}
Zielenbach, W. 2011, \aj, 142, 120

\end{thebibliography}
\bibliographystyle{aa}

\begin{appendix}


\section{Online tables and figures}
\begin{table*}
\caption{\label{tab:ao}Disk-resolved images available for Psyche. Each line gives the epoch, the telescope/instrument, the filter, the exposure time, the airmass, the distance to the Earth $\Delta$ and the Sun $r$, the phase angle $\alpha$, angular size $\phi$ and the PI of the observations.}
\centering
\begin{tabular}{rrl lrr rrrr l}
\hline 
\multicolumn{1}{c} {Date} & \multicolumn{1}{c} {UT} & \multicolumn{1}{c} {Instrument} & \multicolumn{1}{c} {Filter} & \multicolumn{1}{c} {Exp} & \multicolumn{1}{c} {Airmass} & \multicolumn{1}{c} {$\Delta$} & \multicolumn{1}{c} {$r$} & \multicolumn{1}{c} {$\alpha$} & \multicolumn{1}{c} {$\phi$} & PI \\
\multicolumn{1}{c} {} & \multicolumn{1}{c} {} & \multicolumn{1}{c} {} & \multicolumn{1}{c} {} & \multicolumn{1}{c} {(s)} & \multicolumn{1}{c} {} & \multicolumn{1}{c} {(au)} & \multicolumn{1}{c} {(au)} & \multicolumn{1}{c}{(\degr)} & \multicolumn{1}{c} {(arcsec)} &  \\
\hline\hline
  2002-03-06 &     10:16:19 & Keck/NIRC2 &    Kp    &   1 & 1.02 & 2.24 & 3.22 & 1.5  & 0.138 &              Merline \\
  2002-05-08 &      5:50:07 & Keck/NIRC2 &     H    &   4 & 1.01 & 2.82 & 3.27 & 17.1 & 0.110 &               Margot \\
  2003-07-14 &      7:52:19 & Keck/NIRC2 &    Kp    &   3 & 1.44 & 2.72 & 3.20 & 17.4 & 0.114 &              Merline \\
  2009-08-16 &      8:50:07 & Keck/NIRC2 & FeII     &   5 & 1.26 & 1.69 & 2.69 & 4.5  & 0.183 &            Marchis \\
  2010-10-06 &     13:27:56 & Keck/NIRC2 &    Kp    &   0.3 & 1.06 & 2.09 & 2.61 & 21.0 & 0.148 &           Armandroff \\
  2010-10-06 &     14:28:00 & Keck/NIRC2 &     K    &   0.3 & 1.00 & 2.09 & 2.61 & 21.0 & 0.148 &           Armandroff \\
  2010-10-06 &     15:07:06 & Keck/NIRC2 &     H    &   0.3 & 1.00 & 2.08 & 2.61 & 21.0 & 0.148 &           Armandroff \\
  2010-12-13 &      2:11:49 & VLT/NACO   &   Ks &   1 & 1.63 & 1.70 & 2.68 & 2.5 & 0.182 &               Carry \\
  2010-12-13 &      3:10:42 & VLT/NACO   &   Ks &   1 & 1.43 & 1.70 & 2.68 & 2.5 & 0.182 &               Carry \\
  2010-12-13 &      3:54:15 & VLT/NACO   &   Ks &   1 & 1.37 & 1.70 & 2.68 & 2.5 & 0.182 &               Carry \\
  2010-12-13 &      5:21:14 & VLT/NACO   &   Ks &   1 & 1.43 & 1.70 & 2.68 & 2.5 & 0.182 &               Carry \\
  2010-12-14 &      3:58:14 & VLT/NACO   &   Ks &   1 & 1.36 & 1.71 & 2.68 & 2.9 & 0.181 &               Carry \\
  2010-12-14 &      4:04:43 & VLT/NACO   &   Ks &   1 & 1.36 & 1.71 & 2.68 & 2.9 & 0.181 &               Carry \\
  2010-12-14 &      4:42:34 & VLT/NACO   &   Ks &   1 & 1.37 & 1.71 & 2.68 & 2.9 & 0.181 &               Carry \\
  2010-12-14 &      6:11:40 & VLT/NACO   &   Ks   &   1 & 1.62 & 1.71 & 2.68 & 2.9 & 0.181 &               Carry \\
  2015-12-25 &      8:52:42 & Keck/NIRC2 &   Kp   &   0.4 & 1.00 & 1.76 & 2.70 & 7.1  & 0.175 &              de Pater \\
  2015-12-25 &      9:48:43 & Keck/NIRC2 &   Kp   &   0.4 & 1.02 & 1.76 & 2.70 & 7.1  & 0.175 &              de Pater \\
  2015-12-25 &     10:32:56 & Keck/NIRC2 &   Kp   &   0.4 & 1.08 & 1.76 & 2.70 & 7.1  & 0.175 &              de Pater \\
  2015-12-25 &     11:21:28 & Keck/NIRC2 &   Kp   &   0.4 & 1.20 & 1.76 & 2.70 & 7.1  & 0.175 &              de Pater \\
  2018-04-24 &      8:42:52 & VLT/SPHERE & N$\_$R & 245 & 1.31 & 2.29 & 3.26 & 5.8 & 0.135 &               Vernazza \\
  2018-04-24 &      8:47:07 & VLT/SPHERE & N$\_$R & 245 & 1.33 & 2.29 & 3.26 & 5.8 & 0.135 &               Vernazza \\
  2018-04-24 &      8:51:24 & VLT/SPHERE & N$\_$R & 245 & 1.35 & 2.29 & 3.26 & 5.8 & 0.135 &               Vernazza \\
  2018-04-24 &      8:55:38 & VLT/SPHERE & N$\_$R & 245 & 1.37 & 2.29 & 3.26 & 5.8 & 0.135 &               Vernazza \\
  2018-04-24 &      8:59:53 & VLT/SPHERE & N$\_$R & 245 & 1.40 & 2.29 & 3.26 & 5.8 & 0.135 &               Vernazza \\
  2018-04-28 &      7:42:39 & VLT/SPHERE & N$\_$R & 245 & 1.17 & 2.27 & 3.26 & 4.5 & 0.136 &               Vernazza \\
  2018-04-28 &      7:46:54 & VLT/SPHERE & N$\_$R & 245 & 1.18 & 2.27 & 3.26 & 4.5 & 0.136 &               Vernazza \\
  2018-04-28 &      7:51:11 & VLT/SPHERE & N$\_$R & 245 & 1.19 & 2.27 & 3.26 & 4.5 & 0.136 &               Vernazza \\
  2018-04-28 &      7:55:25 & VLT/SPHERE & N$\_$R & 245 & 1.21 & 2.27 & 3.26 & 4.5 & 0.136 &               Vernazza \\
  2018-04-28 &      7:59:39 & VLT/SPHERE & N$\_$R & 245 & 1.22 & 2.27 & 3.26 & 4.5 & 0.136 &               Vernazza \\
  2018-05-04 &      5:44:47 & VLT/SPHERE & N$\_$R & 245 & 1.03 & 2.25 & 3.25 & 2.6 & 0.137 &               Vernazza \\
  2018-05-04 &      5:49:02 & VLT/SPHERE & N$\_$R & 245 & 1.03 & 2.25 & 3.25 & 2.6 & 0.137 &               Vernazza \\
  2018-05-04 &      5:53:17 & VLT/SPHERE & N$\_$R & 245 & 1.03 & 2.25 & 3.25 & 2.6 & 0.137 &               Vernazza \\
  2018-05-04 &      5:57:31 & VLT/SPHERE & N$\_$R & 245 & 1.04 & 2.25 & 3.25 & 2.6 & 0.137 &               Vernazza \\
  2018-05-04 &      6:01:45 & VLT/SPHERE & N$\_$R & 245 & 1.04 & 2.25 & 3.25 & 2.6 & 0.137 &               Vernazza \\
  2018-05-05 &      1:34:27 & VLT/SPHERE & N$\_$R & 245 & 1.58 & 2.25 & 3.25 & 2.3 & 0.137 &               Vernazza \\
  2018-05-05 &      1:38:43 & VLT/SPHERE & N$\_$R & 245 & 1.55 & 2.25 & 3.25 & 2.3 & 0.137 &               Vernazza \\
  2018-05-05 &      1:42:57 & VLT/SPHERE & N$\_$R & 245 & 1.52 & 2.25 & 3.25 & 2.3 & 0.137 &               Vernazza \\
  2018-05-05 &      1:47:11 & VLT/SPHERE & N$\_$R & 245 & 1.49 & 2.25 & 3.25 & 2.3 & 0.137 &               Vernazza \\
  2018-05-05 &      1:51:26 & VLT/SPHERE & N$\_$R & 245 & 1.46 & 2.25 & 3.25 & 2.3 & 0.137 &               Vernazza \\
  2018-06-03 &      3:58:13 & VLT/SPHERE & N$\_$R & 245 & 1.33 & 2.30 & 3.23 & 8.7 & 0.134 &               Vernazza \\
  2018-06-04 &      0:02:27 & VLT/SPHERE & N$\_$R & 245 & 1.31 & 2.30 & 3.23 & 8.7 & 0.134 &               Vernazza \\
  2018-06-04 &      0:06:43 & VLT/SPHERE & N$\_$R & 245 & 1.29 & 2.30 & 3.23 & 8.7 & 0.134 &               Vernazza \\
  2018-06-04 &      0:10:58 & VLT/SPHERE & N$\_$R & 245 & 1.28 & 2.30 & 3.23 & 8.7 & 0.134 &               Vernazza \\
  2018-06-04 &      0:15:11 & VLT/SPHERE & N$\_$R & 245 & 1.26 & 2.30 & 3.23 & 8.7 & 0.134 &               Vernazza \\
\hline
\end{tabular}
\end{table*}

\onecolumn
\begin{longtable}{lr rrr r l}
\caption{\label{tab:lcs}Optical photometry utilized in the shape modeling. The table gives the epoch, the number of individual measurements $N_p$, asteroid's distances to the Earth $\Delta$ and the Sun $r$, phase angle $\varphi$, photometric filter and a reference.}\\
\hline
\multicolumn{1}{c} {Epoch} & \multicolumn{1}{c} {$N_p$} & \multicolumn{1}{c} {$\Delta$} & \multicolumn{1}{c} {$r$} & \multicolumn{1}{c} {$\varphi$} & \multicolumn{1}{c} {Filter}  & Reference \\
 &  &  (au) & (au) & (\degr) &   \\
\hline\hline

\endfirsthead
\caption{continued.}\\

\hline
\multicolumn{1}{c} {Epoch} & \multicolumn{1}{c} {$N_p$} & \multicolumn{1}{c} {$\Delta$} & \multicolumn{1}{c} {$r$} & \multicolumn{1}{c} {$\varphi$} & \multicolumn{1}{c} {Filter}  & Reference \\
&  &  (au) & (au) & (\degr) &   \\
\hline\hline
\endhead
\hline
\endfoot
\hline
1955-12-26.3  &  102  &  1.78  &  2.69  &  9.6   &  V      &    \citet{vanhouten1958}  \\
1956-01-02.3  &  54   &  1.83  &  2.70  &  12.1  &  V      &    \citet{vanhouten1958}  \\
1965-12-05.6  &  32   &  1.67  &  2.65  &  2.1   &  V      &    \citet{chang1963}                           \\
1965-12-18.5  &  24   &  1.71  &  2.66  &  7.0   &  V      &    \citet{chang1963}                           \\
1965-12-19.6  &  65   &  1.72  &  2.66  &  7.4   &  V      &    \citet{chang1963}                           \\
1970-10-26.1  &  90   &  1.82  &  2.61  &  16.1  &  V      &    \citet{Tedesco1985}                    \\
1970-10-27.0  &  82   &  1.81  &  2.61  &  15.8  &  V      &    \citet{Tedesco1985}                    \\
1970-10-29.0  &  94   &  1.80  &  2.61  &  15.2  &  V      &    \citet{Tedesco1985}                    \\
1970-11-04.1  &  83   &  1.75  &  2.61  &  13.2  &  V      &    \citet{Tedesco1985}                    \\
1970-11-23.0  &  92   &  1.67  &  2.64  &  5.7   &  V      &    \citet{Tedesco1985}                    \\
1970-11-25.0  &  94   &  1.67  &  2.64  &  4.9   &  V      &    \citet{Tedesco1985}                    \\
1970-11-27.0  &  131  &  1.66  &  2.64  &  4.0   &  V      &    \citet{Tedesco1985}                    \\
1970-12-03.9  &  26   &  1.66  &  2.65  &  1.8   &  V      &    \citet{Tedesco1985}                    \\
1970-12-05.0  &  94   &  1.67  &  2.65  &  1.8   &  V      &    \citet{Tedesco1985}                    \\
1970-12-07.0  &  40   &  1.67  &  2.65  &  2.1   &  V      &    \citet{Tedesco1985}                    \\
1972-02-06.3  &  29   &  2.29  &  3.20  &  8.3   &  V      &    \citet{taylor1977}                          \\
1972-02-20.0  &  80   &  2.23  &  3.21  &  3.5   &  V      &    \citet{Tedesco1985}                    \\
1972-02-21.0  &  96   &  2.23  &  3.21  &  3.1   &  V      &    \citet{Tedesco1985}                    \\
1972-03-09.0  &  49   &  2.24  &  3.22  &  3.2   &  V      &    \citet{Tedesco1985}                    \\
1973-04-29.0  &  75   &  2.27  &  3.27  &  3.2   &  V      &    \citet{Tedesco1985}                    \\
1973-05-12.4  &  22   &  2.25  &  3.26  &  2.4   &  V      &    \citet{Tedesco1985}                    \\
1973-05-27.3  &  68   &  2.29  &  3.25  &  7.2   &  V      &    \citet{Tedesco1985}                    \\
1974-08-08.4  &  45   &  1.70  &  2.71  &  2.5   &  V      &    \citet{taylor1977}                          \\
1974-08-10.4  &  56   &  1.70  &  2.71  &  3.3   &  V      &    \citet{taylor1977}                          \\
1974-08-11.3  &  161  &  1.70  &  2.71  &  3.7   &  V      &    \citet{taylor1977}                          \\
1974-08-12.3  &  129  &  1.70  &  2.71  &  4.1   &  V      &    \citet{taylor1977}                          \\
1974-08-13.2  &  31   &  1.71  &  2.70  &  4.6   &  V      &    \citet{taylor1977}                          \\
1975-10-03.4  &  55   &  2.07  &  2.59  &  21.2  &  V      &    \citet{Tedesco1983}               \\
1975-10-04.1  &  65   &  2.06  &  2.59  &  21.1  &  V      &    \citet{Tedesco1983}               \\
1975-10-11.4  &  60   &  1.98  &  2.60  &  19.9  &  V      &    \citet{Tedesco1983}               \\
1975-10-13.5  &  24   &  1.96  &  2.60  &  19.5  &  V      &    \citet{Tedesco1983}               \\
1975-10-27.0  &  77   &  1.84  &  2.62  &  16.2  &  V      &    \citet{Tedesco1983}               \\
1975-11-03.3  &  102  &  1.78  &  2.62  &  13.9  &  V      &    \citet{Tedesco1983}               \\
1975-11-08.3  &  118  &  1.75  &  2.63  &  12.2  &  V      &    \citet{Tedesco1983}               \\
1975-11-10.9  &  65   &  1.73  &  2.63  &  11.2  &  V      &    \citet{Tedesco1983}               \\
1975-11-22.9  &  106  &  1.69  &  2.64  &  6.3   &  V      &    \citet{Tedesco1983}               \\
1975-11-23.9  &  76   &  1.68  &  2.65  &  5.9   &  V      &    \citet{Tedesco1983}               \\
1975-12-01.7  &  33   &  1.67  &  2.65  &  2.8   &  V      &    \citet{Tedesco1983}               \\
1975-12-02.7  &  45   &  1.67  &  2.66  &  2.4   &  V      &    \citet{Tedesco1983}               \\
1975-12-04.9  &  53   &  1.68  &  2.66  &  1.9   &  V      &    \citet{Tedesco1983}               \\
1975-12-05.9  &  78   &  1.68  &  2.66  &  1.8   &  V      &    \citet{Tedesco1983}               \\
1975-12-06.3  &  91   &  1.68  &  2.66  &  1.8   &  V      &    \citet{Tedesco1983}               \\
1975-12-06.8  &  117  &  1.68  &  2.66  &  1.8   &  V      &    \citet{Tedesco1983}               \\
1975-12-07.3  &  72   &  1.68  &  2.66  &  1.8   &  V      &    \citet{Tedesco1983}               \\
1975-12-13.7  &  32   &  1.69  &  2.67  &  3.8   &  V      &    \citet{Tedesco1983}               \\
1975-12-14.7  &  63   &  1.70  &  2.67  &  4.1   &  V      &    \citet{Tedesco1983}               \\
1975-12-22.8  &  86   &  1.74  &  2.68  &  7.4   &  V      &    \citet{Tedesco1983}               \\
1975-12-28.8  &  109  &  1.78  &  2.69  &  9.8   &  V      &    \citet{Tedesco1983}               \\
1976-01-04.2  &  41   &  1.83  &  2.70  &  12.0  &  V      &    \citet{Tedesco1983}               \\
1976-01-09.8  &  72   &  1.88  &  2.70  &  13.8  &  V      &    \citet{Tedesco1983}               \\
1976-01-15.8  &  62   &  1.95  &  2.71  &  15.5  &  V      &    \citet{Tedesco1983}               \\
1976-01-22.9  &  24   &  2.03  &  2.72  &  17.2  &  V      &    \citet{Tedesco1983}               \\
1976-03-03.2  &  30   &  2.60  &  2.77  &  20.9  &  V      &    \citet{Tedesco1983}               \\
1977-02-13.4  &  79   &  2.26  &  3.21  &  5.8   &  V      &    \citet{Tedesco1985}                    \\
1978-04-30.8  &  15   &  2.26  &  3.26  &  2.8   &  V      &    \citet{Lupishko1980}                        \\
1978-05-05.8  &  26   &  2.25  &  3.26  &  1.5   &  V      &    \citet{Lupishko1980}                        \\
1978-05-06.8  &  28   &  2.25  &  3.26  &  1.4   &  V      &    \citet{Lupishko1980}                        \\
1978-05-27.8  &  9    &  2.29  &  3.24  &  7.2   &  V      &    \citet{Lupishko1980}                        \\
1978-06-05.7  &  10   &  2.34  &  3.23  &  10.0  &  V      &    \citet{Lupishko1980}                        \\
1979-07-05.9  &  35   &  1.84  &  2.75  &  11.7  &  V      &    \citet{Lupishko1982}                       \\
1979-07-06.9  &  28   &  1.83  &  2.75  &  11.3  &  V      &    \citet{Lupishko1982}                       \\
1979-07-07.9  &  31   &  1.83  &  2.75  &  11.0  &  V      &    \citet{Lupishko1982}                       \\
1979-08-09.8  &  48   &  1.69  &  2.70  &  2.5   &  V      &    \citet{Lupishko1982}                       \\
1979-08-14.8  &  84   &  1.70  &  2.70  &  4.6   &  V      &    \citet{Lupishko1982}                       \\
1980-12-12.7  &  34   &  1.70  &  2.68  &  3.3   &  V      &    \citet{zhou1982}                            \\
1981-01-02.6  &  39   &  1.82  &  2.70  &  11.4  &  V      &    \citet{Zhou1981}                         \\
1981-01-12.6  &  51   &  1.92  &  2.71  &  14.6  &  V      &    \citet{zhou1982}                            \\
1981-04-15.3  &  9    &  3.21  &  2.84  &  17.8  &  V      &    \citet{weidenschilling1987}                 \\
1981-11-04.3  &  7    &  3.41  &  3.11  &  16.6  &  V      &    \citet{weidenschilling1987}                 \\
1981-12-01.3  &  13   &  3.07  &  3.14  &  18.2  &  V      &    \citet{weidenschilling1987}                 \\
1981-12-02.4  &  9    &  3.05  &  3.14  &  18.2  &  V      &    \citet{weidenschilling1987}                 \\
1982-02-04.9  &  45   &  2.31  &  3.20  &  8.8   &  V      &    \citet{Lupishko1982}                        \\
1982-02-17.7  &  102  &  2.25  &  3.21  &  4.4   &  V      &    \citet{Tedesco1985}                    \\
1982-02-18.7  &  71   &  2.25  &  3.22  &  4.1   &  V      &    \citet{Tedesco1985}                    \\
1982-04-24.2  &  7    &  2.64  &  3.26  &  15.5  &  V      &    \citet{harris1999}                      \\
1982-04-24.2  &  7    &  2.64  &  3.26  &  15.5  &  V      &    \citet{harris1999}                      \\
1982-04-25.2  &  5    &  2.66  &  3.26  &  15.6  &  V      &    \citet{harris1999}                      \\
1982-04-25.2  &  5    &  2.66  &  3.26  &  15.6  &  V      &    \citet{harris1999}                      \\
1982-04-28.3  &  6    &  2.70  &  3.27  &  16.1  &  V      &    \citet{harris1999}                      \\
1982-04-28.3  &  6    &  2.70  &  3.27  &  16.1  &  V      &    \citet{harris1999}                      \\
1982-04-29.2  &  13   &  2.71  &  3.27  &  16.2  &  V      &    \citet{harris1999}                      \\
1982-04-29.2  &  13   &  2.71  &  3.27  &  16.2  &  V      &    \citet{harris1999}                      \\
1982-05-21.4  &  18   &  3.02  &  3.28  &  17.9  &  V      &    \citet{weidenschilling1987}                 \\
1982-06-23.4  &  6    &  3.50  &  3.30  &  16.9  &  V      &    \citet{weidenschilling1987}                 \\
1982-06-25.4  &  6    &  3.53  &  3.30  &  16.7  &  V      &    \citet{weidenschilling1987}                 \\
1983-02-20.3  &  18   &  3.00  &  3.30  &  17.3  &  V      &    \citet{weidenschilling1987}                 \\
1983-03-28.4  &  18   &  2.51  &  3.28  &  12.8  &  V      &    \citet{weidenschilling1987}                 \\
1983-03-29.3  &  12   &  2.50  &  3.28  &  12.6  &  V      &    \citet{weidenschilling1987}                 \\
1983-05-22.3  &  16   &  2.26  &  3.24  &  5.1   &  V      &    \citet{weidenschilling1987}                 \\
1983-05-23.2  &  14   &  2.26  &  3.24  &  5.4   &  V      &    \citet{weidenschilling1987}                 \\
1983-06-29.2  &  14   &  2.54  &  3.21  &  15.4  &  V      &    \citet{weidenschilling1987}                 \\
1983-06-30.4  &  10   &  2.55  &  3.21  &  15.5  &  V      &    \citet{weidenschilling1987}                 \\
1984-04-09.3  &  5    &  3.00  &  2.87  &  19.5  &  V      &    \citet{weidenschilling1987}                 \\
1984-05-08.3  &  7    &  2.58  &  2.83  &  20.8  &  V      &    \citet{weidenschilling1987}                 \\
1984-05-09.3  &  11   &  2.56  &  2.83  &  20.8  &  V      &    \citet{weidenschilling1987}                 \\
1984-05-10.3  &  9    &  2.55  &  2.83  &  20.8  &  V      &    \citet{weidenschilling1987}                 \\
1984-05-11.3  &  6    &  2.53  &  2.83  &  20.8  &  V      &    \citet{weidenschilling1987}                 \\
1984-06-07.2  &  10   &  2.16  &  2.79  &  18.5  &  V      &    \citet{weidenschilling1987}                 \\
1984-06-08.4  &  8    &  2.14  &  2.79  &  18.4  &  V      &    \citet{weidenschilling1987}                 \\
1984-06-09.2  &  7    &  2.13  &  2.79  &  18.2  &  V      &    \citet{weidenschilling1987}                 \\
1984-07-25.0  &  11   &  1.73  &  2.73  &  4.4   &  V      &    \citet{Pfleiderer1987}                      \\
1984-07-25.0  &  82   &  1.73  &  2.73  &  4.4   &  V      &    \citet{Pfleiderer1987}                      \\
1984-07-25.1  &  17   &  1.73  &  2.73  &  4.3   &  V      &    \citet{Pfleiderer1987}                      \\
1985-10-25.4  &  25   &  1.86  &  2.63  &  16.5  &  V      &    \citet{weidenschilling1987}                 \\
1985-12-15.0  &  80   &  1.71  &  2.68  &  4.4   &  V      &    \citet{Dotto1992}                           \\
1986-01-19.4  &  34   &  2.00  &  2.73  &  16.4  &  V      &    \citet{weidenschilling1987}                 \\
1987-02-05.3  &  7    &  2.30  &  3.19  &  8.7   &  V      &    \citet{weidenschilling1990}                 \\
1987-02-06.3  &  17   &  2.29  &  3.20  &  8.4   &  V      &    \citet{weidenschilling1990}                 \\
1992-02-17.3  &  53   &  2.24  &  3.20  &  4.8   &  V      &    \citet{neely1992}                                \\
1992-02-22.3  &  25   &  2.23  &  3.21  &  2.9   &  V      &    \citet{neely1992}                                \\
1992-02-23.4  &  22   &  2.23  &  3.21  &  2.5   &  V      &    \citet{neely1992}                                \\
1992-03-06.3  &  44   &  2.23  &  3.22  &  1.9   &  V      &    \citet{neely1992}                                \\
1992-03-12.4  &  47   &  2.25  &  3.22  &  4.2   &  V      &    \citet{neely1992}                                \\
1992-03-13.2  &  23   &  2.25  &  3.22  &  4.5   &  V      &    \citet{neely1992}                                \\
2003-05-28.9  &  24   &  2.29  &  3.24  &  7.1   &  --      &    \citet{Hanus2017b}                              \\
2003-05-29.9  &  36   &  2.29  &  3.24  &  7.4   &  --      &    \citet{Hanus2017b}                              \\
2003-05-31.0  &  23   &  2.30  &  3.24  &  7.8   &  --      &    \citet{Hanus2017b}                         \\
2003-05-32.0  &  31   &  2.30  &  3.24  &  8.1   &  --      &    \citet{Hanus2017b}                         \\
2008-06-23    &  43   &  2.48  &  3.22  &  14.2  &  clear  &    \citep{Grice2017}           \\
2008-06-24    &  103  &  2.49  &  3.22  &  14.4  &  clear  &    \citep{Grice2017}           \\
2008-06-25    &  101  &  2.50  &  3.21  &  14.6  &  clear  &    \citep{Grice2017}           \\
2008-06-29    &  121  &  2.54  &  3.21  &  15.3  &  clear  &    \citep{Grice2017}           \\
2008-06-30    &  123  &  2.55  &  3.21  &  15.5  &  clear  &    \citep{Grice2017}           \\
2008-07-01    &  121  &  2.56  &  3.21  &  15.7  &  clear  &    \citep{Grice2017}           \\
2008-07-02    &  119  &  2.57  &  3.21  &  15.9  &  clear  &    \citep{Grice2017}           \\
2008-07-16    &  147  &  2.74  &  3.20  &  17.7  &  clear  &    \citep{Grice2017}           \\
2008-07-18    &  167  &  2.76  &  3.19  &  17.8  &  clear  &    \citep{Grice2017}           \\
2008-07-19    &  125  &  2.78  &  3.19  &  17.9  &  clear  &    \citep{Grice2017}           \\
2008-07-20    &  118  &  2.79  &  3.19  &  18.0  &  clear  &    \citep{Grice2017}           \\
2008-07-21    &  93   &  2.80  &  3.19  &  18.1  &  clear  &    \citep{Grice2017}           \\
2008-07-22    &  101  &  2.81  &  3.19  &  18.1  &  clear  &    \citep{Grice2017}           \\
2008-07-23    &  107  &  2.83  &  3.19  &  18.2  &  clear  &    \citep{Grice2017}           \\
2008-07-24    &  105  &  2.84  &  3.19  &  18.2  &  clear  &    \citep{Grice2017}           \\
2008-07-25    &  83   &  2.85  &  3.19  &  18.3  &  clear  &    \citep{Grice2017}           \\
2009-05-18    &  90   &  2.43  &  2.81  &  20.7  &  clear  &    \citep{Grice2017}           \\
2009-05-19    &  103  &  2.42  &  2.81  &  20.7  &  clear  &    \citep{Grice2017}           \\
2009-05-22    &  41   &  2.37  &  2.80  &  20.5  &  clear  &    \citep{Grice2017}           \\
2009-05-26    &  121  &  2.32  &  2.80  &  20.2  &  clear  &    \citep{Grice2017}           \\
2009-05-27    &  42   &  2.30  &  2.80  &  20.1  &  clear  &    \citep{Grice2017}           \\
2009-06-01    &  97   &  2.24  &  2.79  &  19.6  &  clear  &    \citep{Grice2017}           \\
2009-06-26    &  115  &  1.93  &  2.75  &  14.9  &  clear  &    \citep{Grice2017}           \\
2009-06-27    &  98   &  1.92  &  2.75  &  14.7  &  clear  &    \citep{Grice2017}           \\
2009-06-29    &  66   &  1.90  &  2.75  &  14.1  &  clear  &    \citep{Grice2017}           \\
2009-06-30    &  107  &  1.89  &  2.75  &  13.8  &  clear  &    \citep{Grice2017}           \\
2009-07-01    &  109  &  1.88  &  2.75  &  13.5  &  clear  &    \citep{Grice2017}           \\
2009-07-02    &  78   &  1.87  &  2.75  &  13.2  &  clear  &    \citep{Grice2017}           \\
2009-07-16    &  95   &  1.76  &  2.73  &  8.4   &  clear  &    \citep{Grice2017}           \\
2009-07-17    &  97   &  1.76  &  2.73  &  8.1   &  clear  &    \citep{Grice2017}           \\
2009-07-18    &  55   &  1.75  &  2.72  &  7.6   &  clear  &    \citep{Grice2017}           \\
2009-07-18    &  97   &  1.75  &  2.73  &  7.7   &  clear  &    \citep{Grice2017}           \\
2009-07-19    &  65   &  1.75  &  2.72  &  7.3   &  clear  &    \citep{Grice2017}           \\
2009-07-19    &  70   &  1.75  &  2.72  &  7.2   &  clear  &    \citep{Grice2017}           \\
2009-07-20    &  62   &  1.74  &  2.72  &  6.9   &  clear  &    \citep{Grice2017}           \\
2009-07-20    &  82   &  1.74  &  2.72  &  6.8   &  clear  &    \citep{Grice2017}           \\
2009-07-21    &  82   &  1.73  &  2.72  &  6.4   &  clear  &    \citep{Grice2017}           \\
2009-07-21    &  98   &  1.74  &  2.72  &  6.5   &  clear  &    \citep{Grice2017}           \\
2009-07-23    &  62   &  1.73  &  2.72  &  5.6   &  clear  &    \citep{Grice2017}           \\
2009-07-24    &  72   &  1.72  &  2.72  &  5.2   &  clear  &    \citep{Grice2017}           \\
2009-07-24    &  77   &  1.72  &  2.72  &  5.3   &  clear  &    \citep{Grice2017}           \\
2009-07-24    &  95   &  1.72  &  2.72  &  5.3   &  clear  &    \citep{Grice2017}           \\
2009-07-25    &  69   &  1.72  &  2.72  &  4.8   &  clear  &    \citep{Grice2017}           \\
2009-07-25    &  95   &  1.72  &  2.72  &  4.9   &  clear  &    \citep{Grice2017}           \\
2009-07-25    &  96   &  1.72  &  2.72  &  4.9   &  clear  &    \citep{Grice2017}           \\
2009-07-26    &  74   &  1.71  &  2.71  &  4.4   &  clear  &    \citep{Grice2017}           \\
2009-07-26    &  107  &  1.71  &  2.71  &  4.4   &  clear  &    \citep{Grice2017}           \\
2009-07-27    &  71   &  1.71  &  2.71  &  4.0   &  clear  &    \citep{Grice2017}           \\
2009-07-28    &  74   &  1.71  &  2.71  &  3.5   &  clear  &    \citep{Grice2017}           \\
2009-07-29    &  70   &  1.70  &  2.71  &  3.1   &  clear  &    \citep{Grice2017}           \\
2009-08-01    &  115  &  1.69  &  2.71  &  1.9   &  clear  &    \citep{Grice2017}           \\
2009-08-01    &  124  &  1.69  &  2.71  &  1.9   &  clear  &    \citep{Grice2017}           \\
2009-08-02    &  51   &  1.69  &  2.71  &  1.5   &  clear  &    \citep{Grice2017}           \\
2009-08-02    &  104  &  1.69  &  2.71  &  1.5   &  clear  &    \citep{Grice2017}           \\
2009-08-09    &  263  &  1.69  &  2.70  &  1.7   &  clear  &    \citep{Grice2017}           \\
2009-08-10    &  91   &  1.69  &  2.70  &  2.1   &  clear  &    \citep{Grice2017}           \\
2009-08-13    &  96   &  1.69  &  2.69  &  3.4   &  clear  &    \citep{Grice2017}           \\
2009-08-15    &  123  &  1.69  &  2.69  &  4.3   &  clear  &    \citep{Grice2017}           \\
2009-08-16    &  65   &  1.69  &  2.69  &  4.7   &  clear  &    \citep{Grice2017}           \\
2009-08-17    &  65   &  1.69  &  2.69  &  5.1   &  clear  &    \citep{Grice2017}           \\
2009-08-20    &  121  &  1.70  &  2.68  &  6.4   &  clear  &    \citep{Grice2017}           \\
2009-08-23    &  121  &  1.71  &  2.68  &  7.6   &  clear  &    \citep{Grice2017}           \\
2009-08-24    &  120  &  1.71  &  2.68  &  8.0   &  clear  &    \citep{Grice2017}           \\
2009-08-27    &  108  &  1.73  &  2.67  &  9.2   &  clear  &    \citep{Grice2017}           \\
2009-09-06    &  138  &  1.78  &  2.66  &  12.9  &  clear  &    \citep{Grice2017}           \\
2009-09-09    &  53   &  1.80  &  2.66  &  13.9  &  clear  &    \citep{Grice2017}           \\
2009-09-10    &  94   &  1.81  &  2.66  &  14.2  &  clear  &    \citep{Grice2017}           \\
2009-09-15    &  61   &  1.85  &  2.65  &  15.7  &  clear  &    \citep{Grice2017}           \\
2009-09-18    &  74   &  1.87  &  2.65  &  16.5  &  clear  &    \citep{Grice2017}           \\
2009-09-22    &  39   &  1.91  &  2.64  &  17.5  &  clear  &    \citep{Grice2017}           \\
2009-10-08    &  40   &  2.08  &  2.63  &  20.6  &  clear  &    \citep{Grice2017}           \\
2010-12-26    &  44   &  1.76  &  2.70  &  7.9   &  clear  &    \citep{Grice2017}           \\
2010-12-30    &  20   &  1.79  &  2.71  &  9.5   &  clear  &    \citep{Grice2017}           \\
2010-12-31    &  85   &  1.80  &  2.71  &  9.8   &  clear  &    \citep{Grice2017}           \\
2010-12-31    &  95   &  1.80  &  2.71  &  9.8   &  clear  &    \citep{Grice2017}           \\
2011-01-01    &  36   &  1.81  &  2.71  &  10.2  &  clear  &    \citep{Grice2017}           \\
2011-01-04    &  53   &  1.83  &  2.71  &  11.2  &  clear  &    \citep{Grice2017}           \\
2011-01-04    &  81   &  1.83  &  2.71  &  11.2  &  clear  &    \citep{Grice2017}           \\
2011-01-06    &  57   &  1.85  &  2.71  &  11.9  &  clear  &    \citep{Grice2017}           \\
2011-01-06    &  85   &  1.85  &  2.71  &  11.9  &  clear  &    \citep{Grice2017}           \\
2011-01-09    &  53   &  1.88  &  2.72  &  12.9  &  clear  &    \citep{Grice2017}           \\
2011-01-09    &  62   &  1.88  &  2.72  &  12.9  &  clear  &    \citep{Grice2017}           \\
2011-01-10    &  68   &  1.88  &  2.72  &  13.2  &  clear  &    \citep{Grice2017}           \\
2011-01-10    &  77   &  1.88  &  2.72  &  13.2  &  clear  &    \citep{Grice2017}           \\
2011-02-03    &  51   &  2.17  &  2.75  &  18.7  &  clear  &    \citep{Grice2017}           \\
2015-11-30  &  238  &  1.70  &  2.67  &  4.6   &  V      &    \citet{Warner2016}                   \\
2015-12-01  &  302  &  1.70  &  2.67  &  4.2   &  V      &    \citet{Warner2016}                   \\
2015-12-26  &  245  &  1.76  &  2.70  &  7.4   &  V      &    \citet{Warner2016}                   \\
2018-9-2.1    &  425  &  3.32  &  3.15  &  17.7  &  Rc   &  Emmanuel Jehin                       \\
2018-8-28.1   &  146  &  3.26  &  3.15  &  18.0  &  Rc   &  Emmanuel Jehin                       \\
2018-8-29.1   &  457  &  3.27  &  3.15  &  18.0  &  Rc   &  Emmanuel Jehin                       \\
\end{longtable}

\begin{table*}[h]
\begin{center}
  \caption[Mass estimates of (16) Psyche]{
    The mass estimates ($\mathcal{M}$) of (16) Psyche collected in the literature.
    For each, the 3\,$\sigma$ uncertainty, 
 computed density (using a diameter of 226.00\,$\pm$\,5.00) and uncertainty,
    method, selection flag, and 
    bibliographic reference are reported. The methods are
    \textsc{defl}: Deflection, \textsc{ephem}: Ephemeris.
    \label{tab:mass}
  }
  \begin{tabular}{rrlclcl}
    \hline\hline
     \multicolumn{1}{c}{\#} & \multicolumn{1}{c}{Mass ($\mathcal{M}$)} &
\multicolumn{1}{c}{$\rho$} & \multicolumn{1}{c}{$\delta \rho$} &
     \multicolumn{1}{c}{Method} & \multicolumn{1}{c}{Sel.} & \multicolumn{1}{c}{Reference}  \\
    & \multicolumn{1}{c}{(kg)} & \multicolumn{1}{c}{(g/cc)} & & & &\\
    \hline
  1 & $(2.53 \pm 1.07) \times 10^{20}$                   &  41.860 &   6.542 & \textsc{defl}  & \ding{55} & \citet{1999-IAA-Vasiliev}                \\
  2 & $(17.30 \pm 15.51) \times 10^{18}$                 &   2.862 &   0.876 & \textsc{defl}  & \ding{51} & \citet{Viateau2000}              \\
  3 & $(4.97 \pm 0.60) \times 10^{19}$                   &   8.223 &   0.637 & \textsc{defl}  & \ding{55} & \citet{2001-IAA-Krasinsky}               \\
  4 & $(6.72 \pm 1.67) \times 10^{19}$                   &  11.118 &   1.181 & \textsc{defl}  & \ding{55} & \citet{Kuzmanoski2002}           \\
  5 & $(2.67 \pm 1.31) \times 10^{19}$                   &   4.418 &   0.782 & \textsc{defl}  & \ding{51} & \citet{2004-SoSyR-38-Kochetova}          \\
  6 & $(2.19 \pm 0.24) \times 10^{19}$                   &   3.623 &   0.274 & \textsc{defl}  & \ding{51} & \citet{2008-DPS-40-Baer}                 \\
  7 & $(79.60 \pm 83.40) \times 10^{18}$                 &  13.170 &   4.682 & \textsc{defl}  & \ding{55} & \citet{2008-PSS-56-Ivantsov}             \\
  8 & $(3.17 \pm 0.19) \times 10^{19}$                   &   5.245 &   0.364 & \textsc{ephem} & \ding{51} & \citet{Fienga2009}               \\
  9 & $(3.35 \pm 1.01) \times 10^{19}$                   &   5.543 &   0.665 & \textsc{ephem} & \ding{51} & \citet{2009-SciNote-Folkner}             \\
 10 & $(4.59 \pm 5.79) \times 10^{19}$                   &   7.594 &   3.233 & \textsc{defl}  & \ding{51} & \citet{Somenzi2010}              \\
 11 & $(3.22 \pm 1.79) \times 10^{19}$                   &   5.328 &   1.049 & \textsc{ephem} & \ding{51} & \citet{2010-IAUS-261-Pitjeva}            \\
 12 & $(2.27 \pm 0.25) \times 10^{19}$                   &   3.756 &   0.285 & \textsc{defl}  & \ding{51} & \citet{2011-AJ-141-Baer}                 \\
 13 & $(24.70 \pm 20.52) \times 10^{18}$                 &   4.087 &   1.164 & \textsc{ephem} & \ding{51} & \citet{2011-Icarus-211-Konopliv}         \\
 14 & $(2.35 \pm 1.18) \times 10^{19}$                   &   3.888 &   0.701 & \textsc{defl}  & \ding{51} & \citet{2011-AJ-142-Zielenbach}           \\
 15 & $(2.46 \pm 0.49) \times 10^{19}$                   &   4.070 &   0.381 & \textsc{defl}  & \ding{51} & \citet{2011-AJ-142-Zielenbach}           \\
 16 & $(2.44 \pm 0.48) \times 10^{19}$                   &   4.037 &   0.378 & \textsc{defl}  & \ding{51} & \citet{2011-AJ-142-Zielenbach}           \\
 17 & $(20.20 \pm 13.02) \times 10^{18}$                 &   3.342 &   0.752 & \textsc{defl}  & \ding{51} & \citet{2011-AJ-142-Zielenbach}           \\
 18 & $(2.51 \pm 1.09) \times 10^{19}$                   &   4.153 &   0.662 & \textsc{ephem} & \ding{51} & \citet{Fienga2011}                  \\
 19 & $(2.51 \pm 1.31) \times 10^{19}$                   &   4.153 &   0.774 & \textsc{ephem} & \ding{51} & \citet{2012-SciNote-Fienga}              \\
 20 & $(17.70 \pm 12.60) \times 10^{18}$                 &   2.929 &   0.722 & \textsc{ephem} & \ding{51} & \citet{2013-Icarus-222-Kuchynka}         \\
 21 & $(2.54 \pm 0.62) \times 10^{19}$                   &   4.203 &   0.439 & \textsc{ephem} & \ding{51} & \citet{2013-SoSyR-47-Pitjeva}            \\
 22 & $(2.23 \pm 1.09) \times 10^{19}$                   &   3.690 &   0.650 & \textsc{ephem} & \ding{51} & \citet{2014-SciNote-Fienga}              \\
 23 & $(2.33 \pm 0.12) \times 10^{19}$                   &   3.855 &   0.264 & \textsc{defl}  & \ding{51} & \citet{2014-AA-565-Goffin}               \\
 24 & $(2.21 \pm 0.16) \times 10^{19}$                   &   3.657 &   0.258 & \textsc{defl}  & \ding{51} & \citet{2014-SoSyR-48-Kochetova}          \\
 25 & $(2.11 \pm 0.64) \times 10^{19}$                   &   3.491 &   0.420 & \textsc{ephem} & \ding{51} & \citet{2017-BDL-108-Viswanathan}         \\
 26 & $(2.29 \pm 0.21) \times 10^{19}$                   &   3.784 &   0.276 & \textsc{ephem} & \ding{51} & \citet{2017-AJ-154-Baer}                 \\
\hline
 & $(2.40 \pm 0.95) \times 10^{19}$ & \multicolumn{2}{c}{Average} \\
   \hline
  \end{tabular}
\end{center}
\end{table*}

\begin{table*}[h]
\begin{center}
  \caption[Diameter estimates of (16) Psyche]{
    The diameter estimates ($\mathcal{D}$) of (16) Psyche collected in the literature.
    For each, the 3\,$\sigma$ uncertainty, method, selection flag, and 
    bibliographic reference are reported. The methods are
    \textsc{adam}: Multidata 3-D Modeling, \textsc{im-te}: Ellipsoid from Imaging, \textsc{lcimg}: 3-D Model scaled with Imaging, \textsc{lcocc}: 3-D Model scaled with Occultation, \textsc{neatm}: Near-Earth Asteroid Thermal Model, \textsc{radar}: Radar Echoes, \textsc{stm}: Standard Thermal Model, \textsc{tpm}: Thermophysical Model.
    \label{tab:diam}
  }
  \begin{tabular}{rrrlclll}
    \hline\hline
     \multicolumn{1}{c}{\#} & \multicolumn{1}{c}{$\mathcal{D}$} & \multicolumn{1}{c}{$\delta \mathcal{D}$} &
\multicolumn{1}{c}{$\rho$} & \multicolumn{1}{c}{$\delta \rho$} &
     \multicolumn{1}{c}{Method} & \multicolumn{1}{c}{Sel.} & \multicolumn{1}{c}{Reference}  \\
    & \multicolumn{1}{c}{(km)} & \multicolumn{1}{c}{(km)} \\
    \hline
  1 &    247.00 &   74.10 &   3.048 &   2.995 & \textsc{stm}   & \ding{55} & \citet{PDSSBN-TRIAD}                     \\
  2 &    253.16 &   12.00 &   2.831 &   1.187 & \textsc{stm}   & \ding{55} & \citet{PDSSBN-IRAS}                      \\
  3 &    262.80 &   12.30 &   2.531 &   1.060 & \textsc{im-te} & \ding{55} & \citet{Drummond2008}         \\
  4 &    222.58 &   16.74 &   4.165 &   1.893 & \textsc{stm}   & \ding{55} & \citet{2010-AJ-140-Ryan}                 \\
  5 &    269.69 &   34.50 &   2.342 &   1.289 & \textsc{neatm} & \ding{55} & \citet{2010-AJ-140-Ryan}                 \\
  6 &    211.00 &   63.00 &   4.889 &   4.786 & \textsc{lcocc} & \ding{55} & \citet{Durech2011}           \\
  7 &    209.00 &   87.00 &   5.031 &   6.589 & \textsc{lcocc} & \ding{55} & \citet{Durech2011}           \\
  8 &    207.22 &    8.94 &   5.162 &   2.144 & \textsc{stm}   & \ding{55} & \citet{Usui2011}                \\
  9 &    244.00 &   24.00 &   3.162 &   1.558 & \textsc{tpm}   & \ding{55} & \citet{2013-Icarus-226-Matter}           \\
 10 &    288.29 &   13.89 &   1.917 &   0.806 & \textsc{neatm} & \ding{55} & \citet{Masiero2012}             \\
 11 &    213.00 &   45.00 &   4.753 &   3.548 & \textsc{lcimg} & \ding{55} & \citet{Hanus2013b}            \\
 12 &    226.00 &   69.00 &   3.979 &   3.968 & \textsc{radar} & \ding{55} & \citet{Shepard2017}             \\
 13 &    225.00 &   12.00 &   4.032 &   1.717 & \textsc{adam}  & \ding{55} & \citet{Hanus2017b}                \\
 14 &    223.00 &   12.00 &   4.142 &   1.766 & \textsc{im-te} & \ding{55} & \citet{Drummond2018}         \\
 15 &    226.00 &   15.00 &   3.979 &   1.759 & \textsc{adam}  & \ding{51} & This work                        \\
\hline
 &    226.00 &     15.00 & \multicolumn{2}{c}{Average} \\
   \hline
  \end{tabular}
\end{center}
\end{table*}

\begin{table*}
\caption{\label{tab:occ}Stellar occultations by Psyche. We list the individual observers.}
\centering
\begin{tabular}{c}
\hline
 Observer \\ \hline\hline
\multicolumn{1}{c} {\textbf{(16) Psyche 2010-08-21}} \\
J. Brooks, Winchester, VA               \\
S. Conard, Gamber, MD                   \\
D. Dunham, Seymour, TX                  \\
A. Scheck, Scaggsville, MD              \\
D. Dunham, Seymour, TX                  \\
D. Dunham, Throckmorton, TX             \\
C. Ellington, Highland Village, TX      \\
P. Maley, Annetta South, TX             \\
R. Tatum, Richmond, VA                  \\
P. Maley, Godley, TX                    \\
H.K. Abramson, Mechanicsville, VA       \\
D. Caton, Boone, NC                     \\
E. Iverson, Athens, TX                  \\
R. Suggs, B. Cooke, Huntsville, AL        \\
J. Faircloth, Kinston, NC               \\
\multicolumn{1}{c} {\textbf{(16) Psyche 2014-07-22}} \\
G. Vaudescal, J. Caquel, R. Yken, J-P. Dupre \\
Jorge Juan, ES                        \\
Carles Schnabel, ES                   \\
C. Perello, A. Selva, ES               \\
Peter Lindner, DE                     \\
Jan Manek, CZ                         \\
Peter Delincak, SK                    \\
Michal Rottenborn, CZ                 \\
\hline
\end{tabular}
\end{table*}

\begin{figure*}
  \begin{center}
    \includegraphics[width=\textwidth]{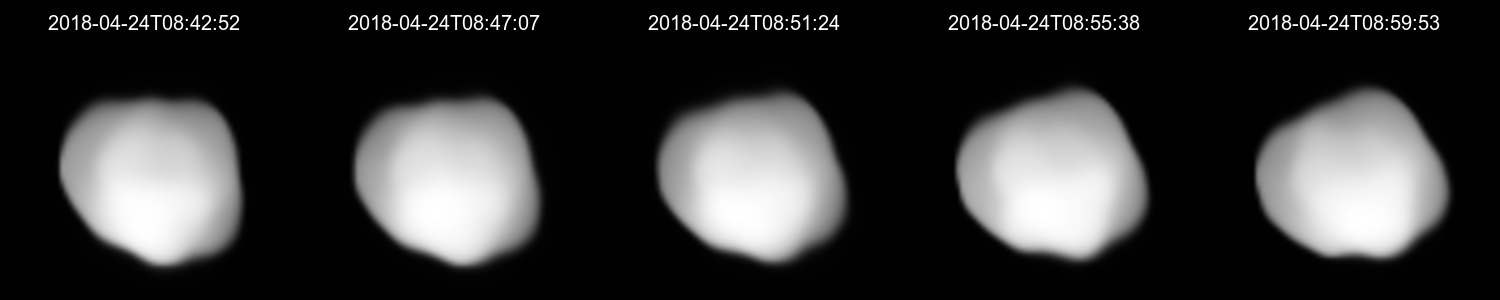}

    \includegraphics[width=\textwidth]{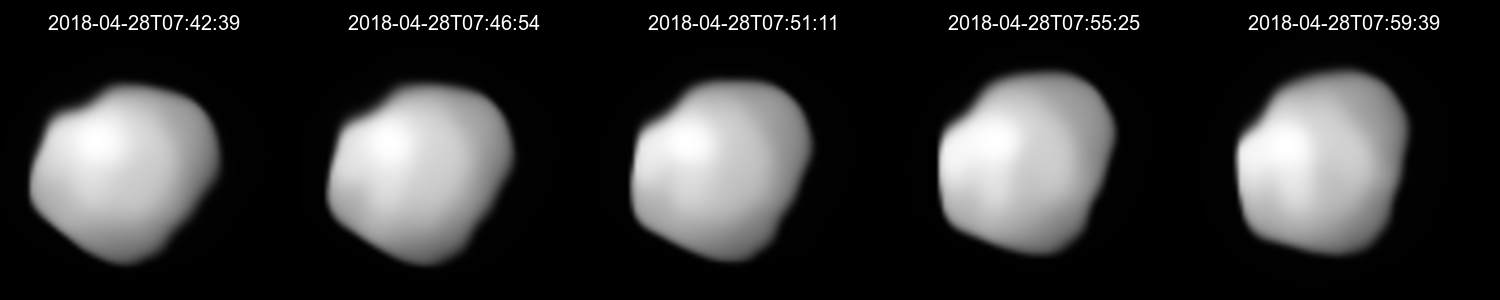}

    \includegraphics[width=\textwidth]{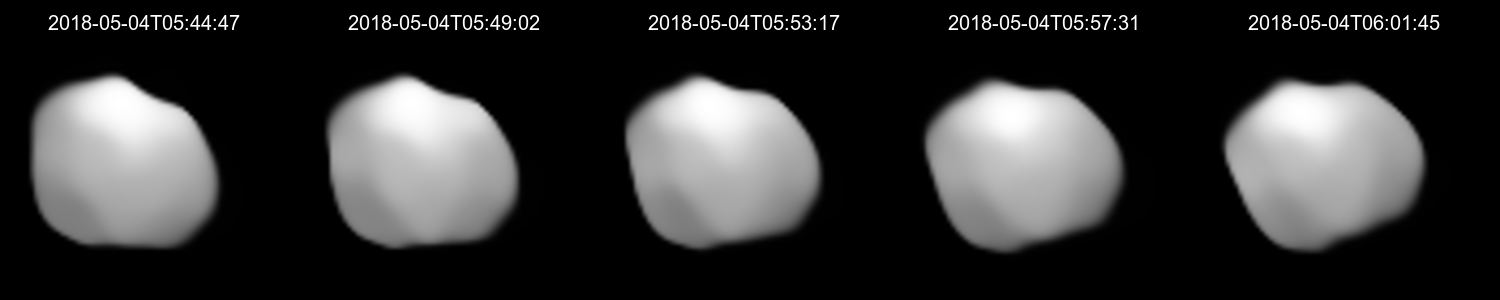}

    \includegraphics[width=\textwidth]{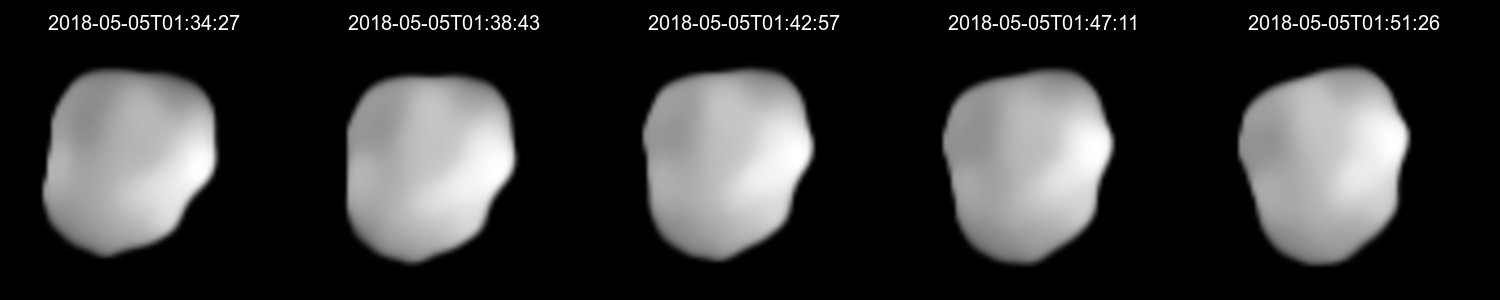}

    \includegraphics[width=\textwidth]{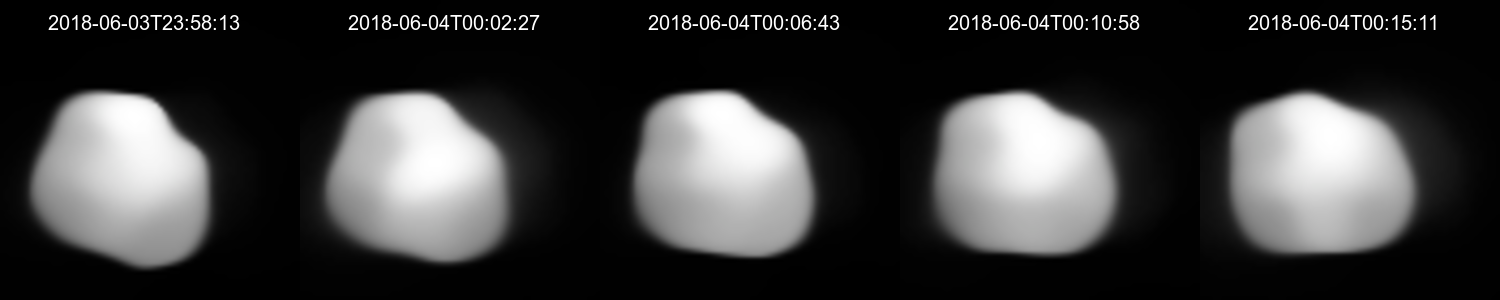}
  \end{center}
  \caption{\label{fig:DeconvAll}All 25 VLT/SPHERE/ZIMPOL images of Psyche obtained at five different epochs and deconvolved with the \mistral{} algorithm and a parametric PSF with a Moffat shape.}
\end{figure*}

\begin{figure*}
  \begin{center}
    \includegraphics[width=0.25\textwidth]{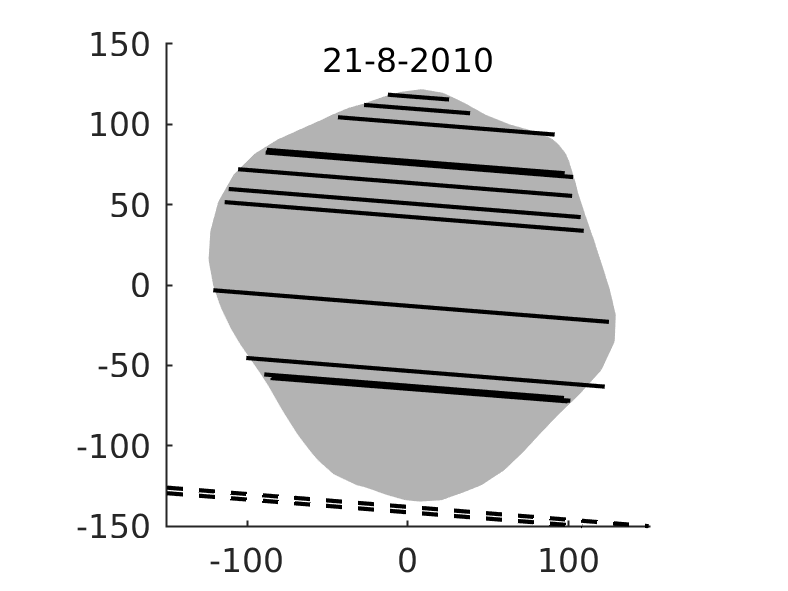}\includegraphics[width=0.25\textwidth]{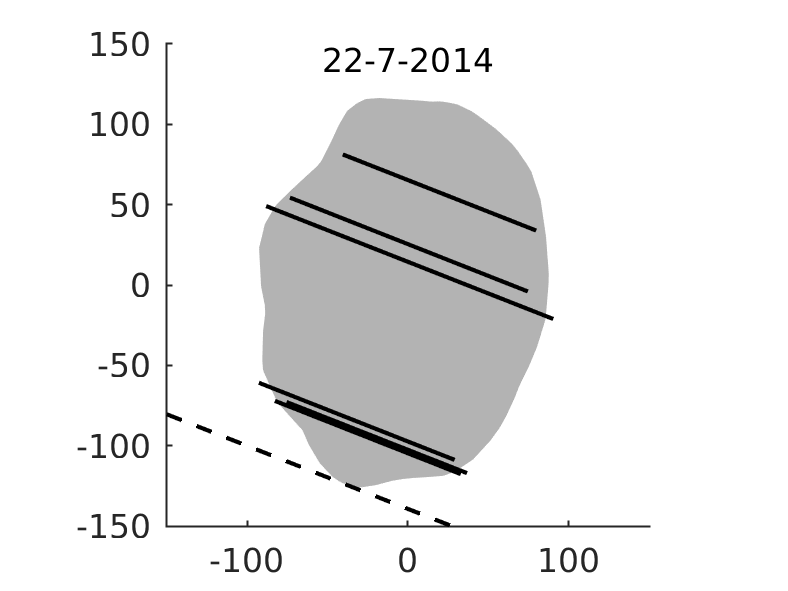}
  \end{center}
  \caption{\label{fig:occ}Observed occultation chords and the  model silhouettes.}
\end{figure*}
\begin{figure*}
\begin{center}
\includegraphics[width=0.32\textwidth]{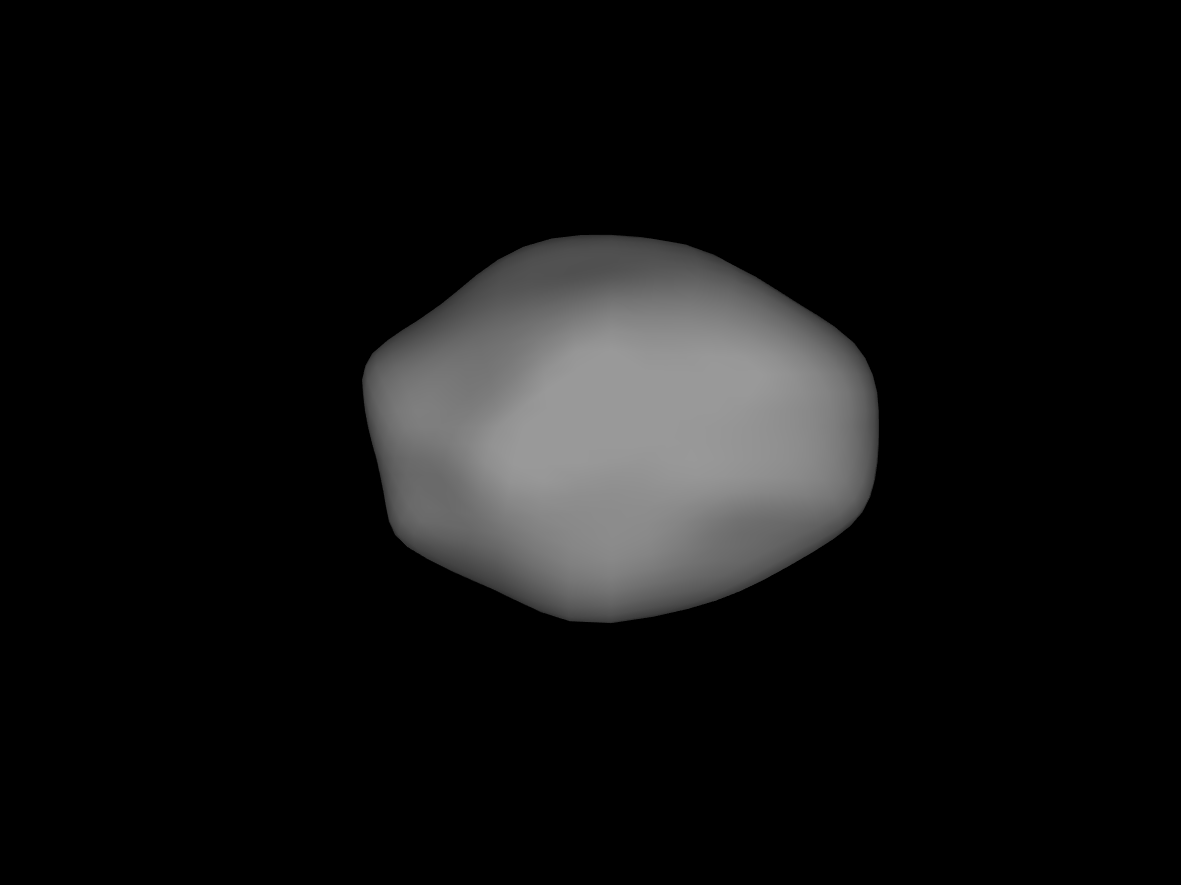}\includegraphics[width=0.32\textwidth]{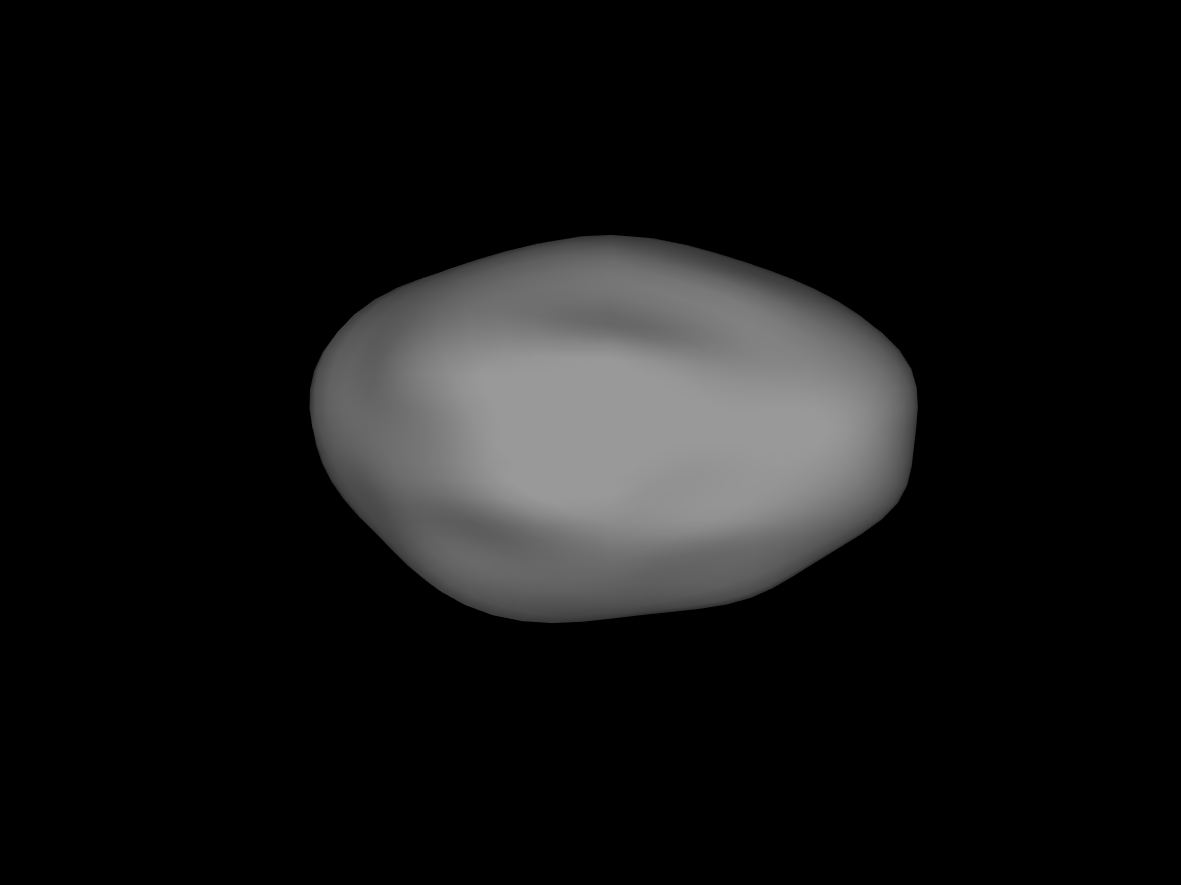}\includegraphics[width=0.32\textwidth]{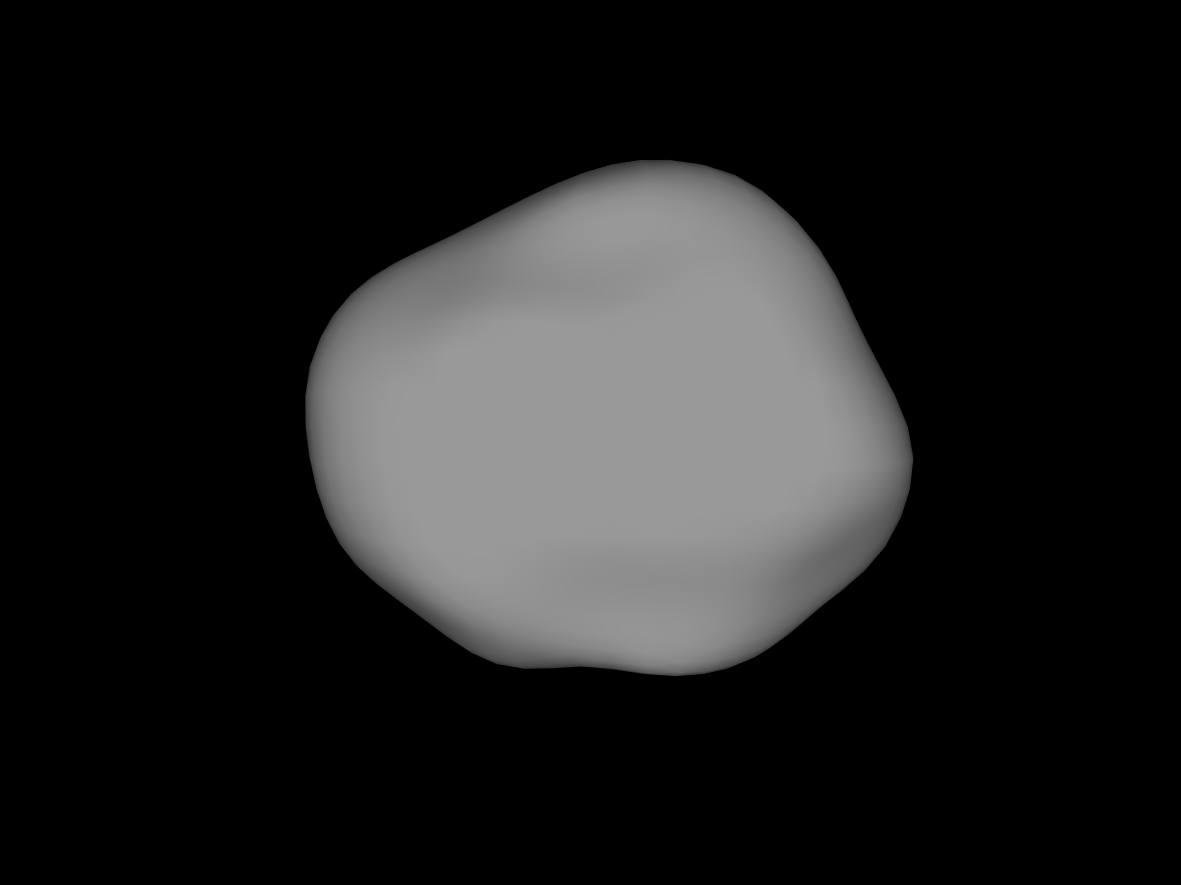}
\end{center}
\caption{\label{fig:model} Shape model of Psyche. Viewing directions are from positive x, y, and z axes, respectively.}
\end{figure*}
\begin{figure*}
  \begin{center}
    \includegraphics[width=0.75\textwidth]{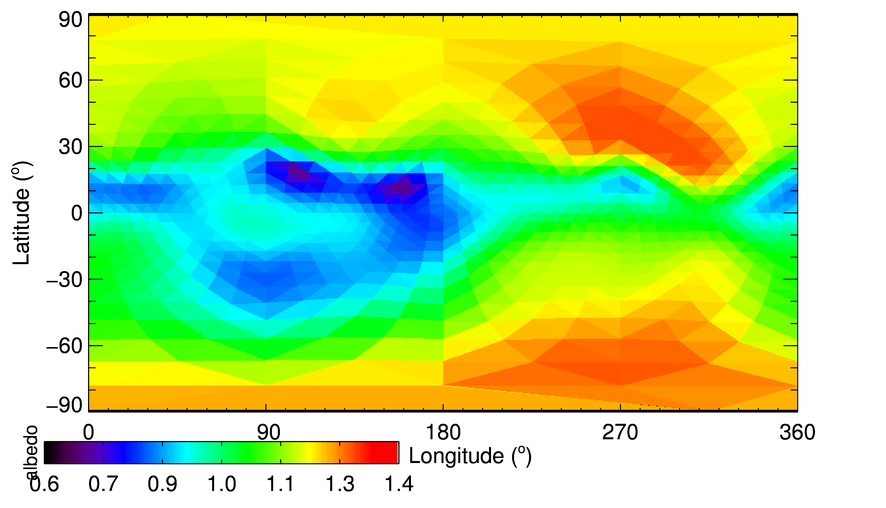}
  \end{center}
  \caption{\label{fig:model_albedo}Albedo map distribution derived from the model showing the presence of a dark region near the equator (\emph{Meroe}) and a bright feature near the poles at a longitude centered to 270 deg.}
\end{figure*}
\begin{figure*}
  \begin{center}
    \includegraphics[width=\textwidth]{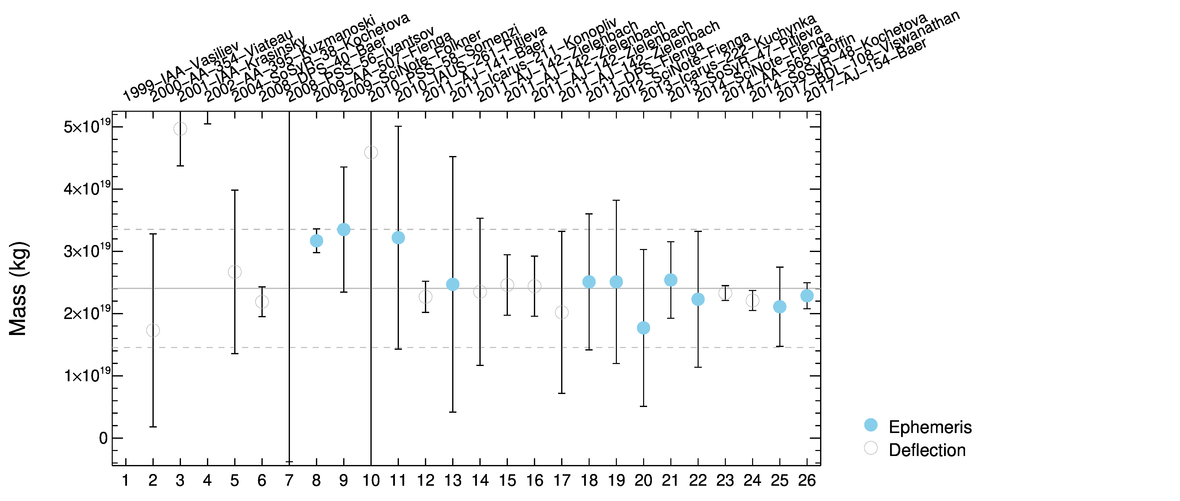}
  \end{center}
  \caption{\label{fig:mass}Available mass estimates for Psyche compiled from the literature. See Table~\ref{tab:mass} for more details.}
\end{figure*}
\begin{figure*}
   \begin{center}
     \includegraphics[width=\textwidth]{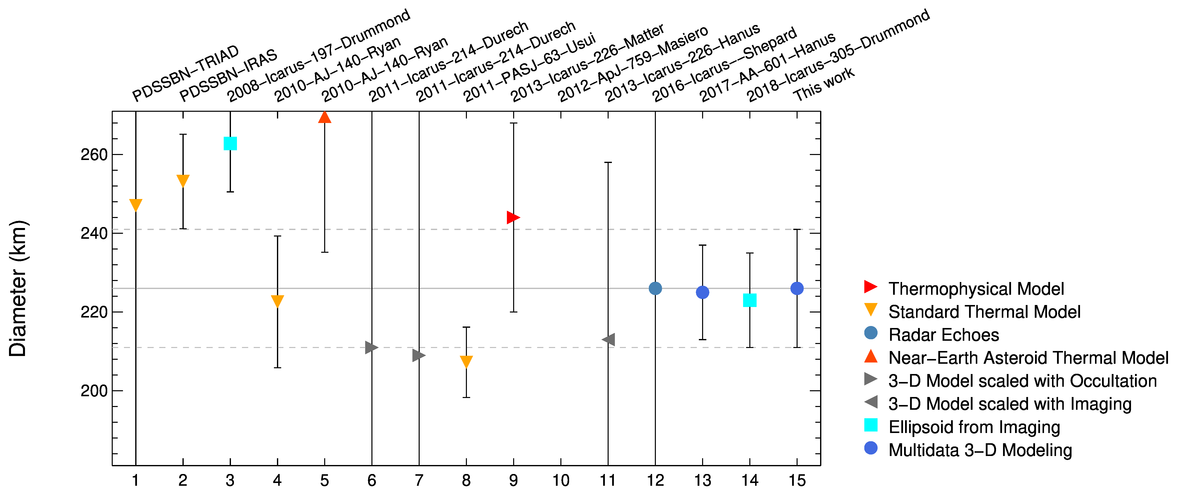}
   \end{center}
   \caption{\label{fig:diam}Available diameter estimates for Psyche compiled from the literature. See Table~\ref{tab:diam} for more details.}
\end{figure*}

\end{appendix}

\end{document}